
\newif\iffigs\figstrue

%
\let\useblackboard=\iftrue
%
%
\newfam\black

\input harvmac.tex

\input epsf

\newcount\figno
\figno=0
\def\fig#1#2#3{
\par\begingroup\parindent=0pt\leftskip=1cm\rightskip=1cm\parindent=0pt
\baselineskip=11pt
\global\advance\figno by 1
\midinsert
\epsfxsize=#3
\centerline{\epsfbox{#2}}
\vskip 12pt
{\bf Fig.\ \the\figno: } #1\par
\endinsert\endgroup\par
}
\def\figlabel#1{\xdef#1{\the\figno}}
\def\encadremath#1{\vbox{\hrule\hbox{\vrule\kern8pt\vbox{\kern8pt
\hbox{$\displaystyle #1$}\kern8pt}
\kern8pt\vrule}\hrule}}
\overfullrule=0pt

\def\Title#1#2{\rightline{#1}
\ifx\answ\bigans\nopagenumbers\pageno0\vskip1in%
\baselineskip 15pt plus 1pt minus 1pt
\else
\def\listrefs{\footatend\vskip 1in\immediate\closeout\rfile\writestoppt
\baselineskip=14pt\centerline{{\bf References}}\bigskip{\frenchspacing%
\parindent=20pt\escapechar=` \input
refs.tmp\vfill\eject}\nonfrenchspacing}
\pageno1\vskip.8in\fi \centerline{\titlefont #2}\vskip .5in}

\ifx\answ\bigans\def\tcbreak#1{}\else\def\tcbreak#1{\cr&{#1}}\fi
\useblackboard
\message{If you do not have msbm (blackboard bold) fonts,}
\message{change the option at the top of the tex file.}
\font\blackboard=msbm10 
\font\blackboards=msbm7
\font\blackboardss=msbm5
\textfont\black=\blackboard
\scriptfont\black=\blackboards
\scriptscriptfont\black=\blackboardss
\def\Bbb#1{{\fam\black\relax#1}}
\else
\def\Bbb#1{{\bf #1}}
\fi
%
\def\yboxit#1#2{\vbox{\hrule height #1 \hbox{\vrule width #1
\vbox{#2}\vrule width #1 }\hrule height #1 }}
\def\fillbox#1{\hbox to #1{\vbox to #1{\vfil}\hfil}}
\def\ybox{{\lower 1.3pt \yboxit{0.4pt}{\fillbox{8pt}}\hskip-0.2pt}}
\def\np#1#2#3{Nucl. Phys. {\bf B#1} (#2) #3}
\def\pl#1#2#3{Phys. Lett. {\bf #1B} (#2) #3}

\def\physrev#1#2#3{Phys. Rev. {\bf D#1} (#2) #3}
\def\anp#1#2#3{Ann. Phys. {\bf #1} (#2) #3}

\def\ijmp#1#2#3{Int. Jour. Mod. Phys. {\bf A#1} (#2) #3}

\def\comments#1{}

\def\Re{{\rm Re\hskip0.1em}}
\def\Im{{\rm Im\hskip0.1em}}

\def\a{\alpha}

\def\II{\relax{I\kern-.07em I}}

\def\IZ{\relax\ifmmode\mathchoice
{\hbox{\cmss Z\kern-.4em Z}}{\hbox{\cmss Z\kern-.4em Z}}
{\lower.9pt\hbox{\cmsss Z\kern-.4em Z}}
{\lower1.2pt\hbox{\cmsss Z\kern-.4em Z}}\else{\cmss Z\kern-.4em
Z}\fi}
\def\IB{\relax{\rm I\kern-.18em B}}
\def\IC{{\bf C}}
\def\ID{\relax{\rm I\kern-.18em D}}
\def\IE{\relax{\rm I\kern-.18em E}}
\def\IF{\relax{\rm I\kern-.18em F}}
\def\IG{\relax\hbox{$\inbar\kern-.3em{\rm G}$}}
\def\IGa{\relax\hbox{${\rm I}\kern-.18em\Gamma$}}
\def\IH{\relax{\rm I\kern-.18em H}}
\def\II{\relax{\rm I\kern-.18em I}}
\def\IK{\relax{\rm I\kern-.18em K}}
\def\IP{\relax{\rm I\kern-.18em P}}

\useblackboard
\def\IZ{\relax\Bbb{Z}}
\fi

\font\cmss=cmss10 \font\cmsss=cmss10 at 7pt
\def\IR{\relax{\rm I\kern-.18em R}}

\def\Re{{\rm Re\ }}
\def\Im{{\rm Im\ }}
\def\BR{\IR}
\def\BZ{\IZ}
\def\BR{\IR}
\def\BC{\IC}


\def\pmod#1{\allowbreak\mkern18mu({mod}\,\,#1)}
\def\log{{{\rm log}\,}}

\def\cos#1{{{\rm cos}{#1}}}
\def\sin#1{{{\rm sin}{#1}}}

\def\lim{{lim}}

\input epsf

\def\SUSY#1{{{\cal N}= {#1}}}                   

\def\wdg{{\wedge}}                              



\def\MR#1{{{\BR}^{#1}}}               
\def\MC#1{{{\BC}^{#1}}}               

\def\MR#1{{{\BR}^{#1}}}               
\def\MC#1{{{\BC}^{#1}}}               
\def\MS#1{{{\bf S}^{#1}}}               
\def\MT#1{{{\bf T}^{#1}}}               

\def\px#1{{\partial_{#1}}}              
\def\qx#1{{\partial^{#1}}}              


\def\Id{{\bf I}}                             

\def\acom#1#2{{\{ {#1},{#2} \}}}               

\def\ev#1{{\langle {#1} \rangle}}           


\def\rep#1{{{\bf {#1}}}}                      



\def\hepth#1{{\it hep-th/{#1}}}

\def\frac#1#2{{{{#1}}\over {{#2}}}}           

\def\u{{\mu}}
\def\v{{\nu}}
\def\b{{\beta}}
\def\g{{\gamma}}

\def\lam{{\lambda}}



\def\Modsp{{\cal M}}     



\def\ut{{\widetilde{u}}}     
\def\ko{{\widetilde{k}}}     
\def\mybar#1{{\overline{{#1}}}}

\def\rad{{\Lambda}}
\def\sc{{\sigma}}
\def\Hv{{\widetilde{H}}}
\def\Bv{{\widetilde{B}}}

\def\wPhi{{\widetilde{\Phi}}}
\def\wrad{{\widetilde{\rad}}}
\def\wsc{{\widetilde{\sc}}} 

\def\wZ{{\widetilde{Z}}}

\def\lieG{{\hat{g}}}




%
\Title{ \vbox{\baselineskip12pt\hbox{hep-th/9812024, PUPT-1792}}}
{\vbox{
\centerline{U-duality Twists and}
\centerline{Possible Phase Transitions in 2+1D Supergravity}
}}
\centerline{
Ori J. Ganor\footnote{$^1$} {origa@puhep1.princeton.edu}
}
\smallskip
\smallskip
\centerline{Department of Physics}
\centerline{Jadwin Hall}
\centerline{Princeton University}
\centerline{Princeton, NJ 08544, USA}
\bigskip
\noindent
We study 2+1D toroidal compactifications of M-theory with twists in the 
U-duality group. These compactifications realize many
symmetric-manifolds from the classification of 2+1D
extended supergravity moduli-spaces.
We then focus on the moduli-space $SU(2,1)/U(2)$ obtained
by dimensional reduction of pure $N=2$ supergravity in 3+1D.
This space is realized with an explicit example.
Assuming that there are no quantum corrections,
we conjecture that the classical discrete duality group has to be
augmented with an extra strong/weak coupling duality.
This implies the existence of new phases of the theory
in which the original 8 compactification radii are all fixed at 
the Planck scale.

\Date{November, 1998}



\nref\rNTW{B. de Wit, A.K. Tollsten and H. Nicolai,
  {``Locally Supersymmetric $D=3$ Nonlinear Sigma-Models,''}
  \np{392}{1993}{3}, \hepth{9208074}}

\nref\rMarSch{N. Marcus and J.H. Schwarz,
  {``Inifinite Symmetry Algebras of Extended Supergravity Theories,''}
  \np{243}{1984}{243}}

\nref\rHulTow{C. M. Hull and P.K. Townsend,
  {``Unity Of Superstring Dualities,''}
  \np{438}{95}{109}, \hepth{9410167}}

\nref\rSenHU{A. Sen,
  {``Strong Weak Coupling Duality in Three Dimensional String Theory,''}
  \np{434}{1995}{179}, \hepth{9408083}}

\nref\rSWII{N. Seiberg and E. Witten,
  {``Monopoles, Duality and Chiral Symmetry Breaking in $N=2$
  Supersymmetric QCD,''} \np{431}{1994}{484--550}, \hepth{9408099}}

\nref\rWitVAR{E. Witten,
  {``String Theory Dynamics in Various Dimensions,''}
  \np{443}{95}{85}, \hepth{9503124}}

\nref\rAPSW{ P. C. Argyres, M. R. Plesser,
  N. Seiberg, E. Witten,
  {``New $N=2$ Superconformal Field Theories In Four Dimensions,''}
  \np{461}{96}{71}, \hepth{9511154}}

\nref\rSeiIRD{N. Seiberg,
  {``IR Dynamics on Branes and Space-Time Geometry,''}
  \pl{384}{1996}{81--85}, \hepth{9606017}}

\nref\rSWGDC{N. Seiberg and E. Witten,
  {``Gauge Dynamics And Compactification To Three Dimensions,''}
  \hepth{9607163}}

\nref\rIntSei{K. Intriligator and N. Seiberg,
  {``Mirror Symmetry in Three Dimensional Gauge Theories,''}
  \pl{387}{1996}{513}, \hepth{9607207}}

\nref\rGMS{O.J. Ganor, D.R. Morrison and N. Seiberg,
  {``Branes, Calabi-Yau Spaces, and Toroidal Compactification
  of the $N=1$ Six-Dimensional $E_8$ Theory,''} 
  \np{487}{1997}{93}, \hepth{9610251}}

\nref\rCFG{S.Cecotti, S.Ferrara and L.Girardello,
  {``Geometry of Type-II Superstrings and
   the Moduli of Superconformal Field Theories,''}
   \ijmp{4}{1989}{2475}}

\nref\rFerSab{S. Ferrara and S. Sabharwal,
  {``Quaternionic Manifolds for type-II
    Superstring Vacua of Calabi-Yau Spaces,''}
    \np{332}{1990}{317}}

\nref\rWGRN{B. de Wit, Grisaru, E. Rabinovici and H. Nicolai,
  \pl{286}{1992}{78}}

\nref\rSenVaf{A. Sen and C. Vafa,
  {``Dual Pairs of Type-II String Compactifications,''}
  \np{455}{1995}{165}, \hepth{9508064}}

\nref\rDasMuk{K. Dasgupta and S. Mukhi,
  {``F-Theory at Constant Coupling,''}
  \pl{385}{1996}{125}, \hepth{9606044}}

\nref\rKumVaf{A. Kumar and C. Vafa,
  {``U-Manifolds,''} \pl{396}{1997}{85-90}, \hepth{9611007}} 

\nref\rDinSil{M. Dine and E. Silverstein,
  {``New M-theory Backgrounds with Frozen Moduli,''}
  \hepth{9712166}}

\nref\rEvaSha{S. Kachru and E. Silverstein,
   {``Selfdual Nonsupersymmetric Typer-II String
   Compactifications,''} \hepth{9808056}}

\nref\rBBS{K. Becker, M. Becker and A. Strominger, unpublished.}

\nref\rStrUH{A. Strominger,
  {``Loop Corrections to the Universal Hypermultiplet,''}
  \pl{421}{1998}{139}, \hepth{9706195}}

\nref\rDabHar{A. Dabholkar and J.A. Harvey,
  {``String Islands,''} \hepth{9809122}}

\nref\rKV{S. Kachru and C. Vafa,
  {``Exact Results For $N=2$ Compactifications
  Of Heterotic Strings,''} \np{450}{95}{69}, \hepth{9505105}}

\nref\rGanZER{O.J. Ganor,
  {``A Note On Zeroes of Superpotentials in F-Theory,''}
  \np{499}{1997}{55}, \hepth{9612077}}

\nref\rIvaVal{E. Ivanov and G. Valent,
  {``Quaternionic Taub-NUT from the harmonic space approach,''}
  \hepth{9809108}}

\nref\rGGV{M.B. Green, M. Gutperle and P. Vanhove,
   {``One loop in eleven dimensions,''} \hepth{9706175}}
\nref\rKPI{E. Kiritsis and B. Pioline,
   {``On $R^4$ threshold corrections in IIB string
   theory and $(p,q)$ string instantons,''} \hepth{9707018}}
\nref\rAPT{I. Antoniadis, B. Pioline and T.R. Taylor,
   {``Calculable $e^{-1/\lambda}$ Effects,''} \hepth{9707222}}
\nref\rGKKOPP{A. Gregori, E. Kiritsis, C. Kounnas, N.A. Obers,
   P.M. Petropoulos, and B. Pioline,
   {``$R^2$ Corrections and Non-perturbative
   Dualities of $N=4$ String ground
   states,''} \np{510}{1998}{423--476}, \hepth{9708062}}
\nref\rKPII{E. Kiritsis and B. Pioline,
  {``U-duality and D-brane Combinatorics,''} \hepth{9710078}}
\nref\rGGI{M.B. Green and M. Gutperle,
   {``$\lambda^{16}$ and related terms in M-theory on $T^2$,''}
   \hepth{9710151}}
\nref\rGGII{M.B. Green  and M. Gutperle,
   {``D-particle bound states and the D-instanton measure,''}
   \hepth{9711107}}
\nref\rOPR{N.A. Obers, B. Pioline and E. Rabinovici,
  {``M-Theory and U-duality on $T^d$ with Gauge Backgrounds,''}
  \hepth{9712084}}
\nref\rP{B. Pioline,
  {``D-effects in Toroidally Compactified Type II String Theory,''}
  \hepth{9712155}}
\nref\rG{M. Gutperle,
   {``Aspects of D-Instantons,''} \hepth{9712156}}
\nref\rSetGre{M.B. Green and S. Sethi,
  {``Supersymmetry Constraints on Type-IIB Supergravity,''}
  \hepth{9808061}}

\nref\rEGKR{S. Elizur, A. Giveon, D. Kutasov and E. Rabinovici,
  {``Algebraic Aspects of Matrix Theory on $T^d$,''}
  \hepth{9707217}}

\nref\rSetSte{S. Sethi and M. Stern,
  {``D-Brane Bound States Redux,''} \hepth{9705046}}
\nref\rYi{P. Yi,
  {``Witten Index and Threshold Bound States of D-Branes,''}
  \hepth{9704098}}
\nref\rPorRoz{M. Porrati and A. Rozenberg,
  {``Bound States at Threshold in
   Supersymmetric Quantum Mechanics,''} \hepth{9708119}}
\nref\rKKN{V.A. Kazakov, I.K. Kostov and N.A. Nekrasov,
  {``D-particles, Matrix Integrals and KP hierachy,''}
  \hepth{9810035}}

\nref\rCJ{E. Cremmer and B. Julia,
  {``The $SO(8)$ Supergravity,''} \np{159}{79}{141}}
\nref\rCJS{E. Cremmer, B. Julia and J. Scherk,
 {``Supergravity Theory in Eleven-Dimensions,''} \pl{76}{78}{409}}

\nref\rGIO{A. Galperin, E. Ivanov and O. Ogievetskii,
  \hepth{9212155},  \anp{230}{1994}{201}}

\nref\rBFSS{T. Banks, W. Fischler, S.H. Shenker and L. Susskind,
  {``M Theory As A Matrix Model: A Conjecture,''}
  \hepth{9610043}, \physrev{55}{1997}{5112-5128}}

\nref\rWitTOP{E. Witten,
  {``Topological Gravity,''} \pl{206}{1988}{601}}
\nref\rPolyak{A.M. Polyakov,
  {``A Few Projects In String Theory,''} \hepth{9304146},
  Published in Les Houches Sum.Sch.1992:0783-804 (QC178:H6:1992)}


\newsec{Introduction}
 In 2+1D the only massless bosonic propagating degrees of freedom
are scalars. The restrictions on the moduli spaces of supergravity
theories in 2+1D have been classified six years ago in \rNTW.
For higher than $\SUSY{4}$ supersymmetries, 
these moduli spaces must be of the form $\Gamma\backslash G/K$,
where $G$ is an appropriate non-compact group, $K$ is its
maximal compact subgroup and $\Gamma$ is a discrete subgroup of $G$.
Compactifications of M-theory provide a concrete realization
of such theories and $\Gamma$ is identified with the group of dualities.
Thus, M-theory on $\MT{8}$ realizes the $\SUSY{16}$
$E_{8(8)}/SO(16)$  moduli space of \rMarSch\ and $E_8(\BZ)$
is the U-duality group \rHulTow.
Similarly, M-theory on $K_3\times \MT{4}$ realizes the
$SO(24,8)/(SO(24)\times SO(8))$ moduli space and $SO(24,8,\BZ)$ is
the group of dualities \rSenHU.

One of the exciting recent developments is the physical interpretation
of certain singularities in the moduli-space
\refs{\rSWII-\rIntSei}.
In general, the assumption of a free theory at low-energies breaks down
when the theory is at a singular point in moduli space.
 A theory with $\SUSY{8}$ rigid supersymmetries in 2+1D can have
a singular point of the form
$\MR{8n}/\Gamma$, where $\Gamma$ is the Weyl-group of a certain
Lie-algebra $\lieG$ of rank $n$. The low-energy description
is then a strongly interacting
conformal field theory. It is defined as the IR limit of 2+1D 
Super-Yang-Mills with the Lie-algebra $\lieG$.
The low-energy description of theories with $\SUSY{4}$ rigid supersymmetry
at singularities is more complicated \refs{\rSeiIRD-\rIntSei}.
For a single vector multiplet it has been classified in \rGMS.
In 3+1D singularities of the moduli space can be connections with  another
phase \rSWII.
In 2+1D, singular points in moduli space of the form $\MR{4}/\Gamma$
are always connections with  another phase (in all known cases).

One of the main motivations for the present work is to explore 
new phases of 2+1D gravity. One direction towards this goal is
to study the singularities of the moduli space of the theory.
The simplest moduli space of an $\SUSY{4}$ supergravity theory
is obtained by dimensional reduction of pure $\SUSY{2}$ supergravity
from 3+1D down to 2+1D. The classical moduli space is
the homogeneous space  $\Gamma_{cl}\backslash SU(2,1)/(SU(2)\times U(1))$
where $\Gamma_{cl}$ is a discrete subgroup
(this moduli-space was introduced in \refs{\rCFG-\rWGRN}).
The classical moduli space
cannot encode the existence of other strongly-coupled phases.
However, we will see that the structure of the moduli space might allow
a consistent extension of $\Gamma_{cl}$ to a U-duality group $\Gamma$.
The extra dualities relate weak coupling to strong coupling and
cannot be seen classically.
The moduli space $\Gamma\backslash SU(2,1)/(SU(2)\times U(1))$
has several singularities of the form $\MR{4}/\Gamma_0$ where $\Gamma_0$
is a finite subgroup of $\Gamma$.
A 2+1D supergravity theory with such a moduli space will most likely have
another phase emanating from these points in moduli space.
In the other phases, the original coupling is stabilized.

To make the discussion more concrete we will construct a particular
compactification of M-theory which realizes the moduli space
$SU(2,1)/(SU(2)\times U(1))$ (at least classically).
In fact, using what we know so far about M-theory, we can realize many
of the other moduli-spaces found in \rNTW\ and hope to study
their discrete duality groups.
The compactification that we will use
is similar in spirit to the compactifications
presented in 
\refs{\rSenVaf-\rEvaSha}
and is obtained in two stages as follows.
We start with a 3+1D vacuum of M-theory with $\SUSY{8}$ supersymmetry.
We can think of it as M-theory on a certain $\MT{7}$, but we pick
a very special $\MT{7}$ (and fluxes) which make it a fixed
point of the U-duality group $E_7(\BZ)$.
Since U-duality is a discrete gauge symmetry, we can now compactify down to
2+1D on $\MS{1}$ with a U-duality twist.
 For an appropriate choice of the twist, we can preserve the desired
amount of supersymmetry.
 The cases in \rNTW\ with more than $\SUSY{4}$ had
$\SUSY{5,6,8,9,10,12,16}$ supersymmetries. We will construct examples
which realize the cases $\SUSY{10,12}$ and some new cases with $\SUSY{8}$.
Given a particular U-duality element $\ut\in E_7(\BZ)$,
we can imagine two ways of 
using it to compactify M-theory down to 2+1D. One way is to mod out by $\ut$
in 3+1D and then compactify on a circle. The second way is to leave
the 3+1D limit intact but compactify on a circle with a twist.
The advantage of the second kind of compactification is that we can be
sure that no extra moduli come from ``twisted sectors.''

As this work was in final stages, I found out that similar ideas
have been put forward in \rBBS\ (referred to in \rStrUH).\foot{I am 
grateful to N. Seiberg for pointing out both these references.}
These papers discuss the universal hyper-multiplet in 3+1D and
the discrete group $\Gamma$ was studied in \rBBS.
Although the setting in the present paper is physically different,
the mathematical details are probably equivalent.

Furthermore, 
recently a paper appeared on the net \rDabHar\ that presents a very
comprehensive study of the moduli spaces of theories obtained via
U-duality twists. 
In \rDabHar\ only $D\ge 4$ dimensions were studied,
but some of the examples presented in sections (2-3) of this paper
are only a special case of \rDabHar.

\newsec{U-duality twists}
The compactifications that we will study are a special 
case of
\refs{\rSenVaf-\rEvaSha}
and are of the following form.
Start with M-theory on $\MT{d}$ which has the moduli space of,
$$
\Modsp_d = E_{d}(\BZ)\backslash E_{d(d)}(\BR) / K_d,
$$
where $K_d$ is the maximal compact subgroup of the non-compact
$E_{d(d)}(\BR)$.
The U-duality group $\Gamma_d\equiv E_{d}(\BZ)$ is, in general, not
a symmetry of the compactification since it maps one point
in $E_{d(d)}(\BR)$ to another. If $\Gamma_d$ were acting freely,
the U-duality group would not have been a symmetry at all. It would
 merely be the first homotopy group of $\Modsp_d$. However, $\Gamma_d$
is not acting freely
 and for a given $\ut\in E_d(\BZ)$, there can be points
in $E_{d(d)}(\BR) / K_d$ which are invariant under $\ut$.
For example, for 9+1D type-IIB compactifications, if $\lam$ is 
the complex dilaton and $\ut$ is the generator of S-duality
which maps $\lam\rightarrow -1/\lam$, then $\lam=i$ is a fixed point of 
$\ut$.
At $\lam=i$ the element $\ut$
is an actual symmetry which acts non-trivially on the states.
Furthermore, it is a gauge symmetry because it can be related to
a geometrical rotation after compactification on a further $\MS{1}$
and mapping type-IIB on $\MS{1}$ to M-theory on $\MT{2}$.
This situation is true in general. An element $\ut\in\Gamma_d$ is
a gauge symmetry when the moduli of M-theory on $\MT{d}$ are derived
from a point in $E_{d(d)}(\BR)/K_d$ which is fixed under $\ut$.

Now suppose we take a point $p\in E_{d(d)}(\BR)/K_d$ which
is invariant under $\ut$ and then take M-theory
on $\MT{d}$ with the moduli corresponding to the point $p$.
Because $\ut$ is a gauge symmetry,
we can compactify on an extra $\MS{1}$ with a twist of $\ut$.
This means that if $\MS{1}$ is very large, after completing a full
revolution around $\MS{1}$ any local state will turn into its $\ut$-dual.

Similarly, we can take a base $\MT{d'}$ and $d'$ U-duality twists
$\ut_1,\dots,\ut_{d'}$ that commute with each other, and then
compactify (M-theory on $\MT{d}$) further on $\MT{d'}$ to get
a compactification down to $(11-d-d')$.
The next task would be to determine how much supersymmetry is left
by such a twist.

\subsec{Supersymmetry}
Part of the U-duality group is actually the geometrical $SL(d,\BZ)$
symmetry. If all the twists are in $SL(d)$ then the compactification
is just M-theory on a $(d+d')$-dimensional manifold and the holonomy
group of the manifold, which is generated by $\ut_1,\dots,\ut_{d'}\in SL(d)$
will determine the amount of supersymmetry.
Sometimes, the group generated by $\ut_1,\dots,\ut_{d'}$
is not in any $SL(d)$ subgroup of $\Gamma_d$,
but after compactification on an auxiliary $\MS{1}$ (unrelated to 
the $\MS{1}$ in the discussion above) and after conjugation by an element
of the $(d+1)$-dimensional U-duality group $\Gamma_{d+1}$, all elements
$\ut_1,\dots,\ut_{d'}$ can be made to lie inside the geometrical $SL(d+1)$.
This statement is actually a special case of the technique 
of \rKumVaf\ used to find generalizations of F-theory.
A simple example of this statement is the U-duality group $SL(5,\BZ)$
for M-theory on $\MT{4}$ which is not entirely geometrical but after
compactification on another $\MS{1}$ it can be conjugated inside
$SO(5,5,\BZ)$ to become the geometrical $SL(5,\BZ)$ of M-theory on $\MT{5}$
(see \rKumVaf\ for further details).

Before we determine 
the amount of supersymmetry left by the twist in the 
general case, let us find the fixed points.
For simplicity we will take $d'=1$.
A fixed point is characterized by elements,
$$
g\in E_{d(d)}(\BR),\qquad \ko\in K_d,
$$
such that
$$
\ut \circ g = g \circ \ko.
$$
To determine the unbroken supersymmetries we need to know how
$\ut$ acts on the spinors. The spinors are in a $2^r$-dimensional
representation of $K_d$.
For $d=8$ they are in the $(\rep{16},\rep{2})$
of $SO(16)\otimes SO(2,1)$. For $d=7$ they are in the
$(\rep{8},\rep{2}) + (\rep{\mybar{8}},\rep{2}')$
of $SU(8)\otimes SO(3,1)$. For $d=6$ they are in the
$(\rep{8},\rep{4})$ of $Sp(4)\otimes SO(4,1)$
(with an extra reality condition since both the $\rep{4}$ of $SO(4,1)$
and $\rep{8}$ of $Sp(4)$ are pseudo-real).
Since $\ut$ is conjugate to $\ko\in K_d$ it is easy to check that
the fraction of unbroken supersymmetry  is $l/2^r$ where $l$ is the
dimension of the eigen-space of $\ko = g^{-1} \ut g$ (in the relevant 
representation of $K_d$) with eigenvalue 1.
 For $d=7$ we have $K_d=SU(8)$. Thus, to preserve at least ${1\over 4}$ of
the supersymmetry, $\ko = g^{-1}\ut g$ must be in an $SU(6)$ subgroup
of $SU(8)$.

\subsec{Counting moduli}
Let us now specialize to the case $d+d'=8$. The compactification
is down to 2+1D and we want to count the number of remaining moduli.
Let us denote by Greek letters ($\u,\v,\dots = 0,1,2$) space-time 
directions, by small English letters ($a,b,\dots = 3,\dots,d'+2$)
the directions of the base $\MT{d'}$ and by capital English
letters ($A,B,\dots = d'+3,\dots 10$) the
local directions of the fiber $\MT{d}$.
Also, $g$ denotes the metric and $C$ the 3-form.
The counting proceeds as follows:
\item{1.}
There are ${{d'(d'+1)}\over 2}$ moduli for the metric $g_{ab}$.
\item{2.}
There are $d'$ moduli for the duals of the vectors $g_{\u a}$.
\item{3.}
There are ${{d'(d'-1)(d'-2)}\over 6}$ moduli for $C_{abc}$.
\item{4.}
There are ${{d'(d'-1)}\over 2}$ moduli for the duals of the vectors
$C_{\u a b}$.
\item{5.}
We need to know $n_s$, the number of scalar moduli of M-theory on $\MT{d}$
that are preserved by the twists.
\item{6.}
We need to know $n_v$, the number of vectors of M-theory on $\MT{d}$
that are preserved by the twists. We then get $(d'+1) n_v$ moduli in 2+1D
if $d'>1$. The extra one is the dual of the vector in 2+1D.
When $d=7$ and $d'=1$
we get just $n_v$ and not $2n_v$, because $n_v$ counts
electric and magnetic duals twice.
\item{7.}
 For $d< 5$ we need to know how many tensors $n_T$ of M-theory
on $\MT{d}$ are preserved. We then obtain ${{d'(d'-1)}\over 2} n_T$
moduli of the form $T_{ab}$ and $d' n_T$ moduli which are duals of
$T_{a\u}$. For $d=5$ and $d'=3$ we get only $3n_T$ because
we count electric-magnetic duals twice.
\item{8.}
We will not consider cases $d\le 3$ here.

Now let us specialize to $d=7$ and $d'=1$.
To count the surviving scalar moduli, $n_s$, we decompose $E_7$ under
$SU(8)\subset E_7$.
The 70 scalar moduli are in the irreducible representation, $\rep{70}$,
of $SU(8)$. This representation is real and
can be represented as the space of the anti-symmetric 4-forms.
We need to know how many eigenvalues of $1$ the element $\ko=g^{-1} \ut g$
possesses in this representation.
To count the vectors $n_v$ we need to know how many eigenvalues
of $1$ appear among the eigenvalues of $\ko$ in the representation,
$\rep{28}+\rep{\mybar{28}}$,
of the vectors.

In the general case of modding out by a U-duality twist,
as studied in \rDinSil, one cannot tell, at present, whether 
there are more moduli that cannot be obtained by simply reducing
the spectrum of M-theory on $\MT{d}$.
For example, if we counted the number of vectors of
M-theory on $\MT{4}/\BZ_2$ by reducing the spectrum of
M-theory on $\MT{4}$ we would not get the 16 blow-up modes.
This $\BZ_2$ was a subgroup of $SL(5,\BZ)$ and it is hard
to tell what happens
for twists which  do not even have a weakly coupled string-theory limit.
On the other hand, in the special cases that we study, one can be
sure that there are no new moduli, since for a large $\MT{d'}$
there is no singularity and the classical approximation is good.
As an example, if we compactified M-theory on $\MT{4}$ further down on
$\MS{1}$ with the same $\BZ_2$ twist as above, we would have got a 
smooth 5-manifold with $SU(2)$ holonomy. It gives a 5+1D vacuum
with $\SUSY{(1,1)}$ supersymmetry and only 8 vectors with an,
$$
SO(4,4,\BZ)\backslash SO(4,4,\BR)/ (SO(4)\times SO(4)),
$$
moduli space. This is in contrast to compactifying M-theory
on $\MT{4}/\BZ_2$ further down on $\MS{1}$ with the,
$$
SO(20,4,\BZ)\backslash SO(20,4,\BR)/ (SO(20)\times SO(4)),
$$
moduli space.



\newsec{Examples}
We now turn to a few examples.
We will start with the cases in 5+1D and then proceed to 2+1D.
Note that only the cases with $\SUSY{8,16}$
in 2+1D can conceivably be (untwisted) toroidal compactifications
of a 3+1D model.

\subsec{Examples in 5+1D}
We start with M-theory on $\MT{4}$ and compactify down on $\MS{1}$
with a U-duality twist that preserves half the supersymmetry.
The U-duality twist is conjugate to an $Sp(2)$ element.
Any element in $Sp(2)$ is conjugate to an element in an $SO(4)$
subgroup. Thus the element $\ut$ is conjugate to a geometrical
twist. This does not necessarily mean that all twists are geometrical
since there might be two elements $\ut_1$ and $\ut_2$ which
are conjugate in $SL(5,\BR)$ but non-conjugate in $SL(5,\BZ)$.

The twist can now be represented as
$e^{i\a}\in U(1)\subset SU(2)\subset SO(4)$.
The particular embedding is such that the fundamental $\rep{4}$
of $SO(4)$ becomes two spinors of $SU(2)$.
The possible values of $\a$ are,
$$
\a = {{\pi}\over 3},{{\pi}\over 2}, {{2\pi}\over 3}.
$$

The possible geometrical twists are thus as follows (up to conjugacy
in $SL(4,\BR)$):
\item{1.}
We can twist by the $\BZ_2$ acting on all 4 directions of a generic
$\MT{4}$.
\item{2.}
We take a special $\MT{4}$ which is a product $\MT{2}\times\MT{2}$
and each $\MT{2}$ has a complex structure which is fixed to
one of the values 
$$
\tau = e^{{\pi i}\over 2}, e^{{\pi i}\over 3}.
$$

We determine the number of vectors in 5+1D as follows.
Let $I,J,K,\dots$ denote directions inside $\MT{4}$ and let $6$
denote the direction of the extra $\MS{1}$. $\u,\v,\dots$
denote 5+1D space-time directions.
The dual of the 3-form of M-theory
and the graviton field $g_{\u 6}$  give rise to two vectors in 5+1D.
The vectors of the form $C_{\u 6 I}$ and the 
form $g_{\u I}$ do not survive the $\ut$-twist. The number of
vectors of the form $C_{\u I J}$ which survive the $\ut$-twist 
is equal to the dimension of the invariant subspace of $\ut$
in $\rep{6}$ of $SO(4)$. This number is $6$ if we have a $\BZ_2$ twist,
and is $4$ for the other cases.
The moduli space is,
$$
\Gamma\backslash SO(4,4,\BR)/(SO(4)\times SO(4)),
$$
in the case of the $\BZ_2$ twist, and is,
$$
\Gamma\backslash SO(4,2,\BR)/ (SO(4)\times SO(2)),
$$
in the case of the special twists.
Compactifying these models to 2+1D gives examples of theories 
with moduli spaces,
$$
\Gamma\backslash SO(8,8,\BR)/(SO(8)\times SO(8)),
$$
and,
$$
\Gamma\backslash SO(8,6,\BR)/(SO(8)\times SO(6)),
$$
respectively.




\subsec{Examples in 2+1D}
We now turn to a few examples in 2+1D.
We start with M-theory on $\MT{7}$ and compactify
on $\MS{1}$ with a twist $\ut\in E_7(\BZ)$.
We saw in the previous section
that to preserve $\SUSY{2r}$ supersymmetries in 2+1D,
we need a twist $\ut$ which is conjugate in $E_{7(7)}(\BR)$
to an element $\ko\subset SU(8)$ with exactly $r$ eigenvalues of $1$.
In the classification of \rNTW\ there are also cases
with $\SUSY{9}$ and $\SUSY{5}$. They cannot be realized
by the kind of compactifications we are studying.
Returning to $\SUSY{2r}$, we will calculate the number of moduli
$k=2 + n_v + n_s$ in each case.
Two moduli are the radius $\rad$ of the circle and the 2+1D dual $\Phi$
of the graviton $g_{\u 3}$ where $3$ is the direction of the circle. 
$n_v$ is the number of 3+1D vectors of the compactification of M-theory
on $\MT{7}$ which are left invariant by $\ut$ and $n_s$ is
the number of scalars which are left invariant.

It is also interesting to check whether the twist can be realized
geometrically or not. If the twist is realized geometrically $\ut$
must be in an $SO(7)\subset SU(8)$ subgroup.
The embedding of $SO(7)$ is such that the fundamental $\rep{8}$
of $SU(8)$ becomes the spinor $\rep{8}$ of $SO(7)$.
We can embed it as $SO(7)\subset SO(8)\subset SU(8)$
where the first embedding is such that the vector $\rep{8}_v$ of $SO(8)$
becomes $\rep{7}+\rep{1}$ of $SO(7)$ and the second embedding is such that
$\rep{8}$ of $SU(8)$ becomes the spinor $\rep{8}_s$ of $SO(8)$.
Given an element $g\in SO(8)$, its matrix in the vector 
representation $\rep{8}_v$ can be brought into a block diagonal
form with four $2\times 2$ blocks of the form,
$$
\pmatrix{
 \cos\a_i & \sin\a_i \cr 
-\sin\a_i & \cos \a_i\cr},\qquad -\pi\le \a_i\le \pi.
$$

The quadruple $(\a_1,\a_2,\a_3,\a_4)$ is determined up to permutation
and up to a change in sign,
$$
(\a_1,\a_2,\a_3,\a_4)\longrightarrow
(\epsilon_1\a_1,\epsilon_2\a_2,\epsilon_3\a_3,\epsilon_4\a_4)
\qquad
\epsilon_i = \pm 1,\qquad\prod\epsilon_i = 1.
$$
The matrix of $g$ in the spinor representation $\rep{8}_s$ has
a similar representation with a quadruple $(\b_1,\dots,\b_4)$
where
\eqn\bfroma{\eqalign{
\b_1 &= {{\a_1 + \a_2 + \a_3 + \a_4} \over 2},\cr
\b_2 &= {{\a_1 + \a_2 - \a_3 - \a_4} \over 2},\cr
\b_3 &= {{\a_1 - \a_2 - \a_3 + \a_4} \over 2},\cr
\b_4 &= {{\a_1 + \a_2 - \a_3 - \a_4} \over 2}.
}}
(modulo $2\pi$).
There is of course the ambiguity of $\pi$ in each phase.
Now, if an element of $SU(8)$ is really an element of $SO(8)$
then all the eigenvalues of the matrix which represents it
in $\rep{8}$ of $SU(8)$ must come in pairs of $e^{\pm i\a_i}$
for some 4 phases $\a_i$.
In addition, if it is an element of $SO(7)$ in the above embedding
then it must be that one of the $\b_i$'s is $0$.
Thus, the condition that a twist $\ut$ is geometrical is
that the eigenvalues of $\ut$, as an $SU(8)$ matrix, are
of the form (for a particular choice of labeling of the eigenvalues), 
$$
e^{\pm i \a_i},\qquad i=1\dots 4,\qquad e^{i\sum_1^4\a_i} = 1.
$$
It is also interesting to check if $\ut$ can be represented
as a T-duality element of type-IIA or type-IIB on $\MT{6}$.
In this case $\ut$ must be in the,
$$
SO(6)\otimes SO(6) \sim SU(4)\otimes SU(4)\subset SU(8),
$$
subgroup of $SU(8)$ which means that it must be possible
to group the 8 eigenvalues in two groups of 4 such
that the product in each group is 1.

Finally, in each case listed below, we will identify the eigenvalues
of $\ut$ as an $SU(8)$ matrix.
We will not attempt to find a $g\in E_{7(7)}(\BR)$ such that
$g u g^{-1}\in E_{7}(\BZ)$ but we will check that the characteristic
polynomial $\det (x I - \ut)$ in the representation,
$\rep{28} + \rep{\mybar{28}}$,
of $SU(8)$ has integer coefficients.

\vskip 0.5cm
{\it Cases with $\SUSY{12}$ supersymmetry:}
\smallskip
$\ut\in SU(8)$ must have 6 eigenvalues of 1.
The other two eigenvalues of $\ko$ must be $e^{\pm i\theta}$, and
$\ko\in U(1)\subset SU(8)$.
The possible values of $\theta$ are determined by the requirement that
in the representation $\rep{56}$ of $SU(8)$, the characteristic
polynomial, $\det (x I - \ko)$, should be integral.
It is easy to calculate,
$$
\det_{\rep{56}} (x I - \ko)
 = (x-1)^{32} (x - e^{i\theta})^{12} (x - e^{-i\theta})^{12}.
$$
This leaves the possibilities,
$$
\theta = {\pi\over 3},\, {\pi\over 2},\, {{2\pi}\over 3},\, \pi.
$$
The number of vectors $n_v$ is determined by decomposing
$\rep{28}+\rep{\mybar{28}}$
of $SU(8)$ under $U(1)\subset SU(8)$,
$$
\rep{28}+\rep{\mybar{28}}
 = 12(\rep{1}_{-1}) + 12(\rep{1}_{+1}) + 32(\rep{1}_0).
$$
which leaves us with $n_v = 32$. This number counts a vector
and its dual separately.
As for $n_s$, we need to decompose the representation
$\rep{70}$ of $SU(8)$ under $U(1)\subset SU(8)$. We find that,
$$
\rep{70} = 30(\rep{1}_0) + 20(\rep{1}_{+1}) + 20(\rep{1}_{-1}).
$$
This leaves us with $n_s = 30$.
Altogether, the moduli space has dimension,
$$
2 + n_s + n_v = 64,
$$
which agrees with the entry in table of \rNTW\ 
for $\SUSY{12}$ (repeated in appendix A).

According to the discussion at the beginning, $\ut$ is not a geometrical
twist, but can be put in the form of an $SO(6,6,\BZ)$ T-duality twist.



\vskip 0.5cm
{\it Cases with $\SUSY{10}$ supersymmetry:}
\smallskip
This time $\ut\in SU(8)$ has to have exactly 5 eigenvalues of 1.
Let the other eigenvalues be
$$
e^{i\a_1},\, e^{i\a_2},\, e^{i\a_3},\qquad \a_1 + \a_2 + \a_3 = 0.
$$
The characteristic polynomial in $\rep{28}+\rep{\mybar{28}}$
of $SU(8)$ is,
\eqn\chpolb{\eqalign{
\det_{\rep{56}} (x I - \ko)
 &= (x-1)^{20} P(x)^6,\cr
P(x) &\equiv
 (x - e^{i\a_1}) (x - e^{-i\a_1})
  (x - e^{i\a_2}) (x - e^{-i\a_2})
  (x - e^{i\a_3}) (x - e^{-i\a_3})
}}

This polynomial has integral coefficients only in the
the following cases,  listed up to an $S_3$
permutation and up to an overall minus sign (replacing $\ko$
by $\ko^{-1}$ which clearly does not make any difference)
\eqn\abpos{\eqalign{
\left({\a_1\over {2\pi}},{\a_2\over {2\pi}},{\a_3\over {2\pi}}\right) 
&=
\left({{1}\over {7}}, {{2}\over {7}}, {{4}\over {7}}\right),\,\,
\left({{1}\over {6}}, {{1}\over {6}}, {{4}\over {6}}\right),\,\,
\left({{1}\over {6}}, {{2}\over {6}}, {{3}\over {6}}\right),\,\,
\left({{1}\over {4}}, {{1}\over {4}}, {{2}\over {4}}\right),\,\,
\left({{1}\over {3}}, {{1}\over {3}}, {{1}\over {3}}\right).
\cr
}}
Since $\ut\in SU(3)$, 
to determine $n_v$ and $n_s$ we have to decompose the appropriate 
representations of $SU(8)$ under $SU(3)\subset SU(8)$.
\eqn\deco{\eqalign{
\rep{28}+\rep{\mybar{28}} &= 20(\rep{1}) + 
   6(\rep{3}) + 6(\rep{\mybar{3}}),
\cr
\rep{70} &= 10(\rep{1}) + 10(\rep{3}) + 10(\rep{\mybar{3}}).
}}
Thus, in all cases above, $n_v = 20$ and $n_s = 10$.
Altogether, $n_v + n_s + 2 = 32$ as expected from the table
in appendix (A).

Again, all of these twists are not geometrical but can be
represented as $SO(6,6,\BZ)$ T-duality twists.

\vskip 0.5cm
{\it Cases with $\SUSY{8}$ supersymmetry:}
\smallskip
We denote the four eigenvalues of $\ko$ which are not 1 by,
$$
e^{i\a_j},\qquad j=1\dots 4,\qquad \sum\a_j = 0\pmod{4}.
$$
The characteristic polynomial in $\rep{28}+\rep{\mybar{28}}$
of $SU(8)$ is,
\eqn\chpolb{\eqalign{
\det_{\rep{56}} (x I - \ko)
 &= (x-1)^{12} P(x)^4 Q(x)^{12},\cr
P(x) &\equiv
\prod_1^4 (x-e^{i\a_j})\prod_1^4 (x-e^{-i\a_j}),\cr
Q(x) &\equiv
\prod_{1\le i<j\le 4} (x-e^{i(\a_i+\a_j)})
}}
It has integral coefficients for,
$$
n=2,3,4,5,6,7,8,9,10,12,14,15,16,18,20,24,30.
$$
\eqn\abpos{\eqalign{
\left({\a_1\over {2\pi}},{\a_2\over {2\pi}},
   {\a_3\over {2\pi}},{\a_4\over {2\pi}}\right) 
&=
\left({{1}\over {2}}, {{1}\over {2}}, 
       {{1}\over {2}}, {{1}\over {2}}\right),\,\,
\left({{1}\over {3}}, {{1}\over {3}}, 
       {{2}\over {3}}, {{2}\over {3}}\right),\,\,
\left({{1}\over {4}}, {{1}\over {4}}, 
       {{1}\over {4}}, {{1}\over {4}}\right),\,\,
\left({{1}\over {4}}, {{1}\over {4}}, 
       {{3}\over {4}}, {{3}\over {4}}\right),\,\,
\left({{1}\over {4}}, {{3}\over {4}}, 
        {{2}\over {4}}, {{2}\over {4}}\right),\,\,
\cr &
\left({{1}\over {5}}, {{2}\over {5}}, 
        {{3}\over {5}}, {{4}\over {5}}\right),\,\,
\left({{1}\over {6}}, {{1}\over {6}}, 
        {{1}\over {6}}, {{3}\over {6}}\right),\,\,
\left({{1}\over {6}}, {{1}\over {6}}, 
        {{2}\over {6}}, {{2}\over {6}}\right),\,\,
\left({{1}\over {6}}, {{1}\over {6}}, 
        {{5}\over {6}}, {{5}\over {6}}\right),\,\,
\left({{1}\over {6}}, {{2}\over {6}}, 
        {{4}\over {6}}, {{5}\over {6}}\right),\,\,
\cr &
\left({{1}\over {6}}, {{3}\over {6}}, 
        {{3}\over {6}}, {{5}\over {6}}\right),\,\,
\left({{1}\over {8}}, {{1}\over {8}}, 
        {{3}\over {8}}, {{3}\over {8}}\right),\,\,
\left({{1}\over {8}}, {{3}\over {8}}, 
        {{5}\over {8}}, {{7}\over {8}}\right),\,\,
\left({{1}\over {9}}, {{2}\over {9}}, 
        {{3}\over {9}}, {{4}\over {9}}\right),\,\,
\left({{1}\over {10}}, {{3}\over {10}}, 
        {{7}\over {10}}, {{9}\over {10}}\right),\,\,
\cr &
\left({{1}\over {12}}, {{5}\over {12}}, 
        {{7}\over {12}}, {{11}\over {12}}\right),\,\,
\left({{2}\over {12}}, {{3}\over {12}}, 
        {{9}\over {12}}, {{10}\over {12}}\right),\,\,
\left({{1}\over {15}}, {{2}\over {15}}, 
        {{4}\over {15}}, {{8}\over {15}}\right),\,\,
\left({{1}\over {15}}, {{2}\over {15}},
         {{4}\over {15}}, {{8}\over {15}}\right),\,\,
\cr &
\left({{1}\over {20}}, {{9}\over {20}},
         {{13}\over {20}}, {{17}\over {20}}\right),\,\,
\left({{1}\over {20}}, {{3}\over {20}},
         {{7}\over {20}}, {{9}\over {20}}\right),\,\,
\left({{1}\over {24}}, {{5}\over {24}},
         {{7}\over {24}}, {{11}\over {24}}\right),\,\,
\cr &
\left({{1}\over {24}}, {{11}\over {24}},
         {{17}\over {24}}, {{19}\over {24}}\right),\,\,
\left({{1}\over {30}}, {{17}\over {30}},
        {{19}\over {30}}, {{23}\over {30}}\right).
\cr
}}
To calculate $n_s$ and $n_v$ we decompose the appropriate 
representations of $SU(8)$ under $SU(4)\subset SU(8)$.
\eqn\decob{\eqalign{
\rep{28}+\rep{\mybar{28}} &= 12(\rep{1}) + 
   4(\rep{4}) + 4(\rep{\mybar{4}})+ 2(\rep{6}),
\cr
\rep{70} &= 2(\rep{1}) + 4(\rep{4}) + 4(\rep{\mybar{4}}) + 6(\rep{6}).
\cr
}}
In all the cases above, there is no vector of 
$\rep{4}$ or $\rep{\mybar{4}}$ which is left invariant.
Let $2l$ be the number of vectors of $\rep{6}$ which are
left invariant under $\ut$. This is given by the number of
pairs of eigenvalues which sum up to zero. In the cases above
there are examples with $l=0,1,2,3$.
Then,
$$
2 + n_s + n_v = 2 + 2 + 12 + 16 l= 16 + 16 l.
$$
Thus, the moduli space is $SO(8,k)/ (SO(8)\otimes SO(k))$ with
$k=2,4,6,8$.
Note that this time some of the $\ut$ listed above
can be represented as purely geometrical twists.
In fact, all the geometrical $\ut$'s can be represented as elements
of $SU(6)$ which means that they can be realized as ordinary
compactifications of 3+1D models with $\SUSY{4}$ supersymmetry on 
$\MS{1}$. This does not mean that all {\it vacua} can be represented
in this way. In principle, there could still be inequivalent elements 
$g\in E_{7(7)}(\BR)$ such that $g^{-1} \ut g\in E_7(\BZ)$ are
not conjugate in $E_7(\BZ)$ (although they are
conjugate in $E_{7(7)}(\BR)$).

There are a few cases which are toroidal compactification
of a 5+1D model with $\SUSY{(1,1)}$ supersymmetry.
This can be obtained by compactifying M-theory on $\MT{4}$
further down on $\MS{1}$ with a U-duality twist which preserves
half the supersymmetry. These are the cases discussed in the first 
subsection.

\vskip 0.5cm
{\it Cases with $\SUSY{6}$ supersymmetry:}
\smallskip
In this case we are looking for 5 eigenvalues,
$$
e^{i\a_i},\qquad \sum_1^5\a_i = 0\pmod{2\pi}.
$$
We will not attempt to exhaust all the cases.
We will just point out that, unlike the previous cases,
for $\SUSY{6}$ there are examples where $\ut$ cannot be conjugated
to a T-duality $SO(6,6,\BZ)$ element.
 For example,
\eqn\abpos{\eqalign{
\left({\a_1\over {2\pi}},{\a_2\over {2\pi}},{\a_3\over {2\pi}},
{\a_4\over {2\pi}},{\a_4\over {2\pi}}\right)
&=
\left({{1}\over {11}}, {{3}\over {11}}, {{4}\over {11}},
 {{5}\over {11}}, {{9}\over {11}}\right),
\cr
}}
We confess to not having checked that there really is a U-duality
element in $E_{7(7)}(\BR)$ which conjugates this $SU(8)$ element
into an $E_7(\BZ)$ element. All we know is that its characteristic
polynomial is integral.



\subsec{Examples with $\SUSY{4}$ in 2+1D}
We now present a few examples with $\SUSY{4}$.

\vskip 0.5cm
{\it 12 moduli:}
\smallskip
Here is a variant of a model that has appeared several times in 
the literature.
Take a base $\MT{2}$ and a fiber of the form $\MT{6} = (\MT{2})^3$
such that each $\MT{2}$ has $\tau = {1\over 2} +{\sqrt{3}\over {2}}i$,
i.e. forms a hexagonal lattice. We also require all the $\MT{2}$'s to 
have equal area.
Let $z_1,z_2,z_3$ be complex coordinates on the three $\MT{2}$'s, with
the identifications,
$$
z_i\sim z_i + 1\sim z_i + {1\over 2} +{\sqrt{3}\over {2}}i.
$$
Let us define the elements $\ut_1, \ut_2\in SU(3)$
as follows.
\eqn\utdefx{\eqalign{
\ut_1 :(z_1, z_2, z_3)\rightarrow
      (\omega z_1, \omega z_2, \omega z_3),\cr
\ut_2 :(z_1, z_2, z_3)\rightarrow  (z_3, z_1, z_2),\cr
}}
where $\omega = e^{{i\pi}\over 3}$.
$\ut_1$ and $\ut_2$ preserve the complex structure and 
the holomorphic 3-form.
Now we compactify on $\MT{2}$ with the twists $\ut_1$ and $\ut_2$
in the fiber $\MT{6}$ as we go along the sides of $\MT{2}$.
The resulting manifold is a 4-fold with $SU(3)$ holonomy.
Compactification of M-theory on this manifold preserves
${1\over 4}$-supersymmetry.
The remaining moduli are listed below.
We denote by $i,j,k$ holomorphic coordinates and by $\bar{i},\bar{j},\dots$
anti-holomorphic coordinates. $a,b$ denote directions on the base
and $\u,\v=0,1,2$ are space-time directions. $C_{ABC}$ denotes the VEV
of the M-theory 3-form. The moduli are:
\item{1.}
The shape and size of the base $\MT{2}$ as well as the dual of the
1-form $C_{ab\u}$
made out of the 3-form of M-theory integrated on $\MT{2}$.
Altogether this gives 4 moduli.
\item{2.}
The volume of $\MT{6}$ which is one modulus.
\item{3.}
The 3-form integrated on the $(3,0)$ and $(0,3)$ cycles,
i.e. $C_{ijk}$ and $C_{\bar{i}\bar{j}\bar{k}}$.
Altogether 2 moduli.
\item{4.}
The moduli of the form $C_{ij\bar{k}}$ and $C_{i\bar{j}\bar{k}}$
get multiplied by $\omega$ and $\omega^{-1}$ respectively
under the operation of $\ut_1$ and therefore become massive.
\item{5.}
The only moduli of the form $C_{i\bar{j}a}$ which are invariant
under $\ut_1$ and $\ut_2$ are,
$$
\sum_{k=1}^3 C_{k\bar{k} a}.
$$
This gives 2 more moduli. We also need to add the dual of 
$\sum_{k=1}^3 C_{k\bar{k} \u}$ to obtain a third modulus.
\item{6.}
The duals of the gravitons $g_{\u a}$ give two more moduli.
Altogether we have 12 moduli.

\vskip 0.5cm
{\it 20 moduli:}
\smallskip
We can pick a particular $\MT{6}$ and an element $\ut\in SU(3)$
such that $\ut^7 = I$ and such that $\ut$ preserves only the volume
of $\MT{6}$.
To describe $\ut$, we start with $\MR{7}$ with coordinates
$(x_1,x_2,\dots,x_7)$ and take $\ut_0\in\BZ_7\subset SO(7)$
to be the cyclic permutation.
Since $\ut_0$ preserves the diagonal direction it is actually in $SO(6)$.
Let $\ut$ be the reduced action of $\ut_0$ on the space $\MR{6}$
that is orthogonal to the diagonal $(1,1,1,1,1,1,1)$.
Its eigenvalues are $e^{{2\pi i k}\over 7}$ for $k=1,\dots 6$.
In the spinor representation $\rep{4}$ of $SO(6)$, its eigenvalues
are $e^{{2\pi i}\over 7}, e^{{4\pi i}\over 7}, e^{{8\pi i}\over 7}$
and 1. It thus preserves one spinor.
To describe $\MT{6}$, take the 7 points in $\MR{7}$,
$$
(1,0,0,0,0,0,0),\, (0,1,0,0,0,0,0),\cdots, (0,0,0,0,0,0,1).
$$
They form a 6-plex which is $\ut_0$ invariant.
Taking one of these as a base, the vectors to the other six
are unit vectors of a lattice in $\MR{6}$. Its unit cell is
the requisite $\MT{6}$.
Now, compactify on a further $\MS{1}$ with a $\ut$ twist.
It is easy to check that we end up with a 3+1D model with
$\SUSY{2}$ supersymmetry. The low-energy description
has 3 vector multiplets, 1 hyper-multiplet and
a gravity multiplet (containing a graviphoton).
The radius of the circle is in a hyper-multiplet so
the metric on the vector-multiplet
moduli space can be calculated for large $R$ and does not get quantum 
corrections \rKV. It would be interesting to find a heterotic dual
to this compactification, perhaps along the lines of \rSenVaf.
After further compactification to 2+1D we find 20 scalar moduli
and $k=5$.

\vskip 0.5cm
{\it 4 moduli:}
\smallskip
We now turn to an example which involves a non-geometrical twist
and has only 4 moduli.
This is the example that we will study in detail later on.
We start with M-theory on $\MT{7}$ and compactify
on $\MS{1}$ with a twist $\ut\in E_7(\BZ)$.
We saw in the previous section
that to preserve 8 supersymmetries, i.e. $\SUSY{4}$ in 2+1D,
we need a twist $\ut$ which is conjugate in $E_{7(7)}(\BR)$
to an element $\ko\in SU(6)\subset SU(8)$.
In this case the number of moduli will be $k=2 + n_v + n_s$.
Two moduli are the radius $\rad$ of the circle and the 2+1D dual $\Phi$
of the graviton $g_{\u 3}$ where ``$3$'' is the direction of the circle. 
$n_v$ is the number of 3+1D vectors of the compactification of M-theory
on $\MT{7}$ which are left invariant by $\ut$ and $n_s$ is
the number of scalars which are left invariant.

We now choose $\ko$ to be the diagonal matrix with eigenvalues,
\eqn\koeig{
(\underbrace{e^{{{i\pi}\over 3}},\dots,
   e^{{{i\pi}\over 3}}}_{6\ {\rm times}}, 1,1).
}
In this way, $\ko$ is in the center $\BZ_6\subset SU(6)$.
The number of vectors $n_v$ is determined by decomposing $\rep{28}$
of $SU(8)$ under $SU(6)$,
$$
\rep{28} = \rep{15} + 2(\rep{6}) + \rep{1}.
$$
Under the $\BZ_6$ center, only $\rep{1}$ is invariant which gives
us (together with $\rep{\mybar{28}}$) $n_v = 2$.
The two vectors are electric-magnetic duals.
As for $n_s$, we need to decompose the representation
$\rep{70}$ of $SU(8)$ under $SU(6)\subset SU(8)$. We find that,
$$
\rep{70} = \rep{15} + \rep{\mybar{15}} + \rep{40}.
$$
here $\rep{40}$ is made of the anti-symmetric 3-forms and is real.
These representations are all charged under $\BZ_6$ and therefore
$n_s = 0$.

 The eigenvalues of $\ut$ cannot be separated into two groups
whose product is $1$. Therefore, the twist is not conjugate
to a T-duality twist. This is also obvious from the fact that
there is no extra modulus which could be a string coupling constant.

It remains to prove that there actually exists an element $\ut\in E_7(\BZ)$
which is conjugate to $\ko$ above.
Let us first note the following.
Denote,
$$
\omega = e^{{i\pi}\over 3},\quad \omega^6 = 1.
$$
In the representation $\rep{56}$, the eigenvalues of $\ko$ are as follows.
$\omega$ appears 12 times, $\omega^{-1}$ appears 12 times,
$\omega^2$ appears 15 times, $\omega^{-2}$ appears 15 times and
$1$ appears twice.
The characteristic polynomial is thus,
\eqn\chpol{\eqalign{
\det (x I - \ko) =&
(x-\omega)^{12} (x-\omega^{-1})^{12} 
(x-\omega^2)^{15} (x-\omega^{-2})^{15} (x-1)^2 \cr
 =& (x^2 - x + 1)^{12} (x^2 + x + 1)^{15} (x-1)^2.
\cr
}}
It has integral coefficients, as a matrix that is conjugate to
a matrix in $E_7(\BZ)$ should have.
Note also that the trace is $(-1)$ which means that $\ut$
cannot be separated into two blocks of $28\times 28$ and it mixes
both electric and magnetic charges.

In appendix (C), we will construct an example of such 
a $\ut$ explicitly.


\newsec{The classical limit}
In the remaining of this paper we will concentrate on the last example
from the previous section. This example was the compactification
of M-theory on a fixed $\MT{7}$ further down on $\MS{1}$ of radius
$\rad$ with a U-duality twist that fixes the moduli of $\MT{7}$.
Most of the time,
the particular details of the twist will not concern us.
The moduli space is a quaternionic manifold of dimension 4.
We will assume below that it is
$\Gamma\backslash SU(2,1)/(SU(2)\times U(1))$.
Locally, this is the same moduli space that one obtains from dimensional
reduction of $\SUSY{2}$ 3+1D supergravity (see \refs{\rCFG-\rWGRN}).
To see why we could get the same moduli space in our case
we will first study the classical limit of
large radius $\rad$.
In this limit, we can first take the low-energy limit of M-theory
on $\MT{7}$ at the particular fixed moduli, and then compactify
that 3+1D low-energy action further down on $\MS{1}$ with a twist.
This process involves a subtlety which we shall now discuss.

\subsec{S-duality twists}
The low-energy of M-theory on $\MT{7}$ has 56 electric and magnetic
field-strengths. We have seen that the twist along $\MS{1}$ is not
a mere geometrical transformation (otherwise, the volume of $\MT{7}$
would remain a modulus) and it takes some of the charges
to their magnetic duals.
How do we twist an abelian $U(1)$ theory by S-duality or, more generally,
by the $SL(2,\BZ)$ group?
This topic was discussed in \rGanZER\ in the context of the D3-brane
partition function in F-theory.
The result is as follows. Let $0\le \xi\le 2\pi$ be the coordinate
on $\MS{1}$. To describe the twist,
we must allow the gauge field $A_\u(x,\xi)$ to be 
discontinuous at $\xi=0$.  We take the action,
$$
{1\over {4g^2}}\int d\xi\, d^3x\,F^2,
$$
in the bulk, and add to it a 2+1D Chern-Simons-like interaction,
$$
{1\over {2c}}\int d^3 x \lbrack
a A\wdg dA +  2 A\wdg dA' -f A'\wdg dA'
\rbrack.
$$
Here,
$$
A_\mu(x) \equiv A_\u(x,0),\qquad A'_\u(x)\equiv A_\u(x,2\pi),
$$
and $\pmatrix{a & b\cr c & f\cr}$ is the $SL(2,\BZ)$ duality matrix.
In general, the extra Chern-Simons interaction will make the
gauge field massive.
The only massless gauge fields which remain are those which are
invariant under the twist.

\subsec{Classical Kaluza-Klein reduction}
The action of the remaining zero-modes
in 2+1D is determined, in the limit $\rad\rightarrow\infty$,
by dimensionally reducing the classical action of 3+1D.
The 3+1D action contains 70 scalar fields which, after the twist,
all become massive. It also contains 56 $U(1)$ gauge-fields and
their magnetic duals and
we have seen that only 2 of them are invariant under the $\ut$-twist.
These two must therefore be electric-magnetic duals of each other.
To get the 2+1D dimensionally reduced action, it is thus enough
to start with a 3+1D action of the form,
$$
\int\,d^3 x\,d\xi\,\sqrt{g} \lbrack R(g)
+ {1\over {4 e^2}} \Hv_{\u\v} \Hv^{\u\v}
+{\theta\over {32\pi^2}}\epsilon^{\u\v\b\g}\Hv_{\u\v} \Hv_{\b\g}
 \rbrack.
$$
Here $\xi$ is the coordinate along $\MS{1}$,
$g_{\u\v}$ is the 3+1D metric, $\Hv$ is the field-strength of the
$U(1)$ gauge field $\Bv_\u$
and the coupling constant is $e$.
The coupling constant can be determined
from the fixed point $g$ of $\ut$ in
the moduli space of M-theory on $\MT{7}$.
Since I do not know what $g$ is (in principle there 
might even be several nonequivalent solutions) we will keep it
as a parameter.

Let us recall the procedure for dimensional reduction to 2+1D.
(see \rMarSch).
After reduction to 2+1D and Weyl-rescaling we obtain a Lagrangian of
the form,
\eqn\dimred{\eqalign{
L =& \int\sqrt{g} \left\{
R + {2\over {\rad^2}}\px{\u}\rad\qx{\u}\rad
    + \pi^2\rad^4 F_{\u\v} F^{\u\v}
+ {1\over {2 e^2\rad^2}}\px{\u}\Phi\qx{\u}\Phi
\right\}
\cr &
+\int\sqrt{g} \left\{
 {{\pi^2\rad^2}\over {e^2}} H_{\u\v} H^{\u\v}
+{{\pi^2\rad^2}\over {e^2}}\Phi^2 F_{\u\v} F^{\u\v}
+ {{2\pi^2\rad^2}\over {e^2}} \Phi H_{\u\v} F^{\u\v}
\right\}
\cr &
+{{\theta}\over {16\pi}}\int \epsilon^{\u\v\a} \left[H_{\u\v}\px{\a}\Phi
  + \Phi F_{\u\v}\px{\a}\Phi\right].
\cr
}}
Here,
$$
F_{\u\v} = \px{\u}A_\v - \px{\v} A_\u,
$$
and $A_\u$ is the gauge field coming from the 3+1D graviton.
$\Phi$ is the scalar coming from the 3+1D gauge field $\Bv_\u$,
and $B_\u \equiv \Bv_\u - \Phi A_\u$ is the 2+1D gauge field.
$\rad$ is the radius of $\MS{1}$ (in 3+1D units).
We dualize the two gauge fields $A_\u$ and $B_\u$
into two scalars with integral periods. After a change of variables
we find a sigma-model with metric,
\eqn\prscmet{
ds^2 = {{1}\over {e^4\rad^4}} (d\sc_1^2 + 2 \Phi d\sc_1 d\sc_2 
   + (e^2\rad^2 + \Phi^2) d\sc_2^2) + 4 {{d\rad^2}\over {\rad^2}}
 + {1\over {e^2\rad^2}}d\Phi^2.
}
With the definition,
\eqn\taudef{
\tau = \tau_1 + i\tau_2 = {{2\pi i}\over {e^2}} + {{\theta}\over {2\pi}}.
}
The moduli space is subject to the identifications,
\eqn\clgam{\eqalign{
\sc_1 & \rightarrow\sc_1 - n_3\sc_2 + \tau_2^{-1} n_1 
       - {1\over 2}\tau_2^{-1}\tau_1 n_3^2,
\cr
\sc_2 &\rightarrow\sc_2 + \tau_2^{-1} n_2 
     + \tau_2^{-1}\tau_1 n_3,
\cr
\Phi &\rightarrow\Phi + n_3,
\cr
&\qquad n_1,n_2,n_3\in\BZ.\cr
}}

The two coordinates $d\sc_1-d\sc_2$ form a torus $T^{2}$
with complex structure,
$$
\tau = ie \rad - \Phi,
$$
and  area,
$$
A = {e\over {\rad^3}}.
$$
The target-space metric is Einstein, as required by supersymmetry \rNTW,
and satisfies,
$$
R_{ab} = -{3\over 2} g_{ab},\qquad R = -6.
$$



\subsec{The symmetric space $SU(2,1)/(SU(2)\times U(1))$}
Ignoring the discrete identifications \clgam, for the moment,
 the moduli space
and metric given by \prscmet\ can be identified with the symmetric
K\"ahler manifold $SU(2,1)/(SU(2)\times U(1))$ (see \refs{\rCFG-\rWGRN}).
Let us sketch how this works. The exact details are given 
in appendix (D).

$SU(2,1)$ can be realized as the subgroup of complex $3\times 3$
matrices $M$ with unit determinant, $SL(3,\BC)$, that preserve
an indefinite form which we take to be,
$$
J = \pmatrix{
 0 & 0 & -1 \cr
 0 & 2 &  0 \cr
-1 & 0 &  0 \cr}.
$$
The matrices $M$ satisfy $M^\dagger J M = J$.
We can think of the space $SU(2,1)/(SU(2)\times U(1))$
as the space of complex directions
inside $\MC{3}$ with the indefinite metric,
$$
ds^2 = -dz_1 d\bar{z}_3  -dz_3 d\bar{z}_1 + 2 |dz_2|^2.
$$
We can parameterize the bulk of $SU(2,1)/(SU(2)\times U(1))$
with two complex  coordinates $z_1,z_2$ with a restriction,
\eqn\restzz{
\Re z_1 > |z_2|^2.
}
The K\"ahler function takes the Fubini-Study form,
\eqn\kfunc{
K = \log (z_1 + \bar{z}_1 - 2 |z_2|^2).
}
 The action of an element,
$$
g \equiv \pmatrix{
g_{11} & g_{12} & g_{13} \cr
g_{21} & g_{22} & g_{23} \cr
g_{31} & g_{32} & g_{33} \cr
}\in SU(2,1),
$$
takes the form,
\eqn\gact{
g : (z_1, z_2)\rightarrow
\left(
{{g_{11} z_1 + g_{12} z_2 + g_{13}}
       \over {g_{31} z_1 + g_{32} z_2 + g_{33}}},
{{g_{21} z_1 + g_{22} z_2 + g_{23}}
       \over {g_{31} z_1 + g_{32} z_2 + g_{33}}}
\right).
}
The precise mapping between the variables $(\sc_1,\sc_2,\Phi,\rad)$
and $(z_1,z_2)$ is written down in appendix (D). Here we will
list a few simple consequences. We need to transform
the identifications \clgam\ to the $(z_1,z_2)$ variables.
An identification of the form,
$$
(\sc_1,\sc_2,\Phi,\rad)\sim
(\sc_1+a,\sc_2,\Phi,\rad),
$$
becomes,
$$
(z_1,z_2)\sim (z_1 +4i a, z_2).
$$
Similarly, an identification of the form,
$$
(\sc_1,\sc_2,\Phi,\rad)\sim
(\sc_1,\sc_2+a,\Phi,\rad),
$$
becomes,
$$
(z_1,z_2)\sim (z_1 - 2 a z_2 +a^2, z_2 -a),
$$
and an identification of the form,
$$
(\sc_1,\sc_2,\Phi,\rad)\sim
(\sc_1-a \sc_2,\sc_2,\Phi + a,\rad),
$$
becomes,
$$
(z_1,z_2)\sim (z_1 + 2 i a z_2 + a^2, z_2 -i a).
$$
Thus, using \clgam, the classical moduli space can be taken as
the space of $(z_1,z_2)\in\MC{2}$ subject to \restzz\ and,
$$
0\le\Im z_2 < 1,\qquad
0\le\Re z_2 < \tau_2^{-1},\qquad
0\le\Im z_1 < 4\tau_2^{-1}.
$$
It is also important to know what is the region $\rad\rightarrow\infty$
in terms of $(z_1,z_2)$. As we show in appendix (D), this is the  region,
$$
 \Re z_1 - |z_2|^2 \rightarrow \infty.
$$
The region $(\Re z_1 - |z_2|^2) \rightarrow 0$ is the boundary
of the moduli space and is inaccessible to classical analysis.



\newsec{The quantum moduli space}
We have seen that the moduli of dimensionally reduced 3+1D $\SUSY{2}$
supergravity is given by,
$$
\Modsp_{cl} = \Gamma_{cl}\backslash SU(2,1)/U(2).
$$
$\Gamma_{cl}$ is a discrete group generated by the shifts,
\eqn\disshif{\eqalign{
(z_1,z_2) &\sim (z_1 +4i\tau_2^{-1}, z_2),\cr
(z_1,z_2) &\sim
    (z_1 + 2 \tau_2^{-1} z_2 +\tau_2^{-2}, z_2 +\tau_2^{-1}),\cr
(z_1,z_2) &\sim (z_1 + 2 \tau_2^{-1}\mybar{\tau} z_2
     + \tau_2^{-2} |\tau|^2,
     z_2 + \tau_2^{-1} \tau).\cr
}}
As mentioned above, $\tau$ could be determined if we knew
the moduli at the fixed point.
After a redefinition $z_2\rightarrow \tau_2 z_2$ and
$z_1\rightarrow |\tau_2|^2 z_1$ we find 
the generators of these shifts \disshif\ in $SU(2,1)$ to be,
\eqn\gensh{\eqalign{
g_1 = \pmatrix{
1 & 0 & 4i\tau_2 \cr 
0 & 1 &  0  \cr
0 & 0 &  1  \cr},\qquad
g_2 = \pmatrix{
1 & 2 & 1  \cr 
0 & 1 & 1  \cr
0 & 0 & 1  \cr},\qquad
g_3 = \pmatrix{
1 & 2\mybar{\tau} & |\tau|^2  \cr 
0 & 1 & \tau  \cr
0 & 0 & 1  \cr},
}}
Note that,
$$
g_1 = g_2^{-1}  g_3^{-1}  g_2 g_3
$$
Note also that under $\tau\rightarrow -1/\tau$ the classical group is
invariant provided we conjugate,
$$
g \rightarrow P g P^{-1},
$$
with,
$$
P = \pmatrix{
-{1\over {\tau}} & 0 & 0\cr
0  & 1 & 0 \cr
0 & 0 & -\mybar{\tau} \cr}.
$$

What is the quantum moduli space of the theory?
In principle there could be two kinds of changes from the classical
limit to the full quantum result:
\item{1.}
The group $\Gamma_{cl}$ could be a subgroup of a larger group 
$\Gamma$ of dualities.
\item{2.}
The metric could be corrected, since we only have $\SUSY{4}$ supersymmetry.




\subsec{Strong coupling}
The classical moduli space $SU(2,1)/U(2)$ has a nasty 
infinite volume region
as $\rad\rightarrow 0$. It is unlikely that such a singularity
remains in the quantum theory. There are two conceivable ways in
which this singularity is resolved. 
It could be that quantum corrections smooth out the moduli-space
and in the quantum moduli-space only the classical region
$\rad\rightarrow\infty$ is non-compact.
The other possibility (which is my bet) is that the classical
identification group $\Gamma_{cl}$ is extended to a quantum duality
group $\Gamma$.
What are the restrictions on $\Gamma$? 
Obviously, $\Gamma$ must not contain any element that relates
two classical-vacua, that is vacua with large $\rad$, that are unrelated
by $\Gamma_{cl}$. 
As is usually the case with strong/weak dualities,
the extra generators of $\Gamma$ must take
the weakly-coupled regime $\rad\rightarrow\infty$ to a strongly
coupled regime of $\rad \sim 1$ or $\rad \rightarrow 0$.

As an example, let us consider the element with the matrix,
$$
S = \Omega \pmatrix{
0 & 0 &  1 \cr
0 &-1 &  0 \cr
1 & 0 &  0 \cr} \Omega^{-1}.
$$
For some appropriate $\Omega$. Let us also set $\Omega = \Id$.
it is easy to check that $S$ maps the classical
region $\rad\rightarrow\infty$ to  $\rad\ll 1$.
For $\sc_1=\sc_2=0$ it acts as,
\eqn\radinv{
\Phi + 2i e\rad \longrightarrow -{{1}\over {\Phi + 2i e\rad}}.
}
Let $\Gamma'$ be the group generated by $S$ together with
$\Gamma_{cl}$.
Then, 
$$
\Gamma'\backslash SU(2,1)/U(2),
$$
is a good candidate for the quantum moduli space (for an appropriate 
$\Omega$).

We will discuss the two possibilities, quantum corrections and an 
extended duality group, more in the following subsections.



\subsec{Sources for the quantum corrections}
The first possibility to explore is that quantum effects
correct the metric.
It is not completely clear to me whether mathematically
such perturbation of the Quaternionic-K\"ahler structure is allowed.
What we are looking for is a metric that preserves the
behaviour at $\rad\rightarrow\infty$ while at the same time does not
have singularities which are more nasty than ADE, and the
portion of the moduli space with bounded $\rad$ is compact.
In \rStrUH\ it was shown that
the contribution of quantum corrections to the universal
hypermultiplet metric in type-II compactifications on a Calabi-Yau
can be absorbed in a redefinition of the variables.
On the mathematical side, techniques for obtaining quaternionic-K\"ahler
metrics have been developed in \refs{\rGIO,\rIvaVal}.\foot{I am
grateful to A. Strominger for pointing out these references.} It has
been shown there that the metric can be encoded in a single
analytic function (denoted by ${\cal L}^{+4}$).
We will not attempt to study the possible deformations to the
metric in this paper. Below, we will list a few possible sources
for quantum corrections.
My bet is that these quantum corrections vanish.

\vskip 0.5cm
{\it From $\lam^{16}$ terms in 3+1D:}
\smallskip
Toroidal compactifications
of M-theory have calculable $R^4$ corrections
\refs{\rGGV-\rSetGre}. They are accompanied by 16-fermion terms
with a calculable coefficient \rGGI.
In particular, M-theory on $\MT{7}$ has such a 16-fermion term, although
it has not been calculated explicitly.
Unless  the coefficient vanishes at the particular point
in moduli space in which we are working, it could correct the metric 
as follows.
Once we compactify on $\MS{1}$, most of the fermions
will not be invariant under the twist $\ut$ and will acquire a mass
of order $1/\rad$ in 3+1D Einstein units.
In 2+1D, after Weyl-rescaling the mass becomes $1/\rad^2$.
$\SUSY{4}$ supersymmetry in 2+1D permits
4-fermion terms. Starting with a $\lam^{16}$ term in 3+1D
we get a $\lam^{4}$ term if 12 fermions get a mass.


\vskip 0.5cm
{\it From charged particles in 3+1D:}
\smallskip
In principle we can get terms which behave as 
$e^{-m\rad-iQ\Phi}$ from particles with mass $m$ and
charge $Q$ under the surviving $U(1)$ gauge field in 3+1D.
We can get an instanton by letting the particle have a Euclidean
world-line around $\MS{1}$.
There are also similar terms from the monopoles.
In appendix (B) we show that
such a particle is ${1\over 8}$-BPS and not ${1\over 2}$-BPS.
Such a particle has 28 zero-modes corresponding to the broken
supersymmetries. We will show in appendix (B) that 4 of the zero
modes are invariant under $\ut$ while the other $24$ transform
with a phase and hence get a mass of the order of ${1\over\rad}$
after the compactification on $\MS{1}$.
If there were just one particle with the minimum charge we would
conclude that the quantum corrections are nonzero.
However, ${1\over 8}$-BPS states usually have a large multiplicity.
For example, to calculate the multiplicity of
bound states of D2-branes on $\MT{6}$ which form
a holomorphic curve inside $\MT{6}$ one has to calculate the
cohomology $H^*({\cal M})$ of the moduli space of such curves.
The net contribution to the instanton term at lowest order would be
proportional to the Euler number $\chi({\cal M})$. Usually,
the moduli space ${\cal M}$ is singular and one has to resolve it
before calculating the Euler number. In the case of curves in $\MT{4}$
the Euler number turns out to be zero. It is quite likely that
the net contribution is zero in our case as well.

\vskip 0.5cm
{\it From Kaluza-Klein monopoles:}
\smallskip
We could also make a 2+1D instanton
out of a KK monopole with respect to $\MS{1}$ which is ``wrapped''
over the volume of $\MT{7}$.
If there were no twist, such an instanton would definitely exist
and its contribution would have been proportional to
$e^{-c R^2 V + i\sigma_2}$.
Here, $V$ is the volume of $\MT{7}$, in 3+1D Einstein units.
(This factor can be calculated from the tension of a D6-brane.)
However, it is not clear to me what happens to the instanton
after the twist. for large $\rad$, the KK monopole geometry is smooth
and has a small curvature. Since our twist $\ut$
was an element of $\BZ_6$ we can build our vacuum by modding out 
of M-theory on $\MT{7}\times \widetilde{\MS{1}}$ where $\widetilde{\MS{1}}$
has a radius of $6\rad$.
Now we can actually construct a KK monopole solution for 
$\MT{7}\times\widetilde{\MS{1}}$ and mod that solution out by 
$\BZ_6$.  If $\BZ_6$ were acting freely, everything would have been fine.
The problem is that $\BZ_6$ has a fixed point at the center of the 
KK monopole solution. At that fixed point, the action is a combination
of the geometrical $\BZ_6$ and a non-geometrical $\ut$ and it is not
clear how to study it. (See \rDinSil\ for an attempt to use M(atrix)
theory \rBFSS\ to study such cases.)


\subsec{Extended group of dualities}
We will now discuss in more detail the possibility that the moduli space
is,
$$
\Gamma\backslash SU(2,1)/U(2),
$$
where $\Gamma\supset\Gamma_{cl}$ is a quantum group of dualities.
We must first argue that there is a way to extend $\Gamma_{cl}$
without making any unwanted classical identification and
so that the dangerous region $\rad\rightarrow 0$ is mapped
to the classical region.
We will not give a general proof, but consider the following
argument.
Suppose we start with M-theory on $\MT{7}\times \MS{1}$.
We can embed the U-duality group $E_7(\BZ)$
of M-theory on $\MT{7}$ inside the U-duality group
of M-theory on $\MT{8}$. We can then find the subgroup
$\Gamma'$ of the U-duality group $E_8(\BZ)$ which commutes
with the element $\ut$ of the twist.\foot{I have benefited from
a discussion with N. Seiberg on this point.}
This will be the group that preserves the form $\MT{7}\times\MS{1}$
with $\MT{7}$ fixed at the special moduli.
It is also clear that $\Gamma'$ acts on $SU(2,1)/U(2)$
which is the moduli space of compactifications of the form
$\MT{7}\times \MS{1}$ with $\MT{7}$ fixed. $\Gamma'$ also
contains $\Gamma_{cl}$ and satisfies the requirement about not 
making any unwanted classical identifications.
Thus, if $\Gamma'$ has elements which
map the region $\rad\rightarrow 0$ to $\rad\rightarrow\infty$ it
is a good candidate for $\Gamma$.
Let us show that this is plausible.
It is easier to analyze $E_8(\BZ)$ if we restrict to $\MT{8}$'s
without any fluxes and with right-angles.
The subgroup of $E_8(\BZ)$ which preserves these constraints turns
out to be $W$, the Weyl group of $E_8$ (see \rEGKR).
Let the radii be $R_1,\dots, R_8$.
Then $W$ is generated by permutations in $S_8$ and by
the ``T-duality'' transformation which takes the radii to,
$$
V_{123}^{-2/3} R_1,\,
V_{123}^{-2/3} R_2,\,
V_{123}^{-2/3} R_3,\,
V_{123}^{1/3} R_4,\,
\cdots
V_{123}^{1/3} R_8,\qquad V_{123}\equiv R_1 R_2 R_3.
$$
Let us now perform T-duality on directions $(123)$,
then T-duality on directions $(456)$,
then T-duality on $(678)$,
then T-duality on $(123)$,
then T-duality on $(145)$,
and T-duality on $(678)$.
We finish with a permutation replacing 6 with 1, to obtain the radii,
\eqn\ttxz{\eqalign{
&
r_1 = R_1^{-1} R_2^{-2/3}R_3^{-2/3}R_4^{-1}
                   R_5^{-1}R_6^{-1/3}R_7^{-2/3}R_8^{-2/3},\,
r_2 = R_2^{1/3}R_3^{-2/3}R_6^{-1/3}R_7^{1/3}R_8^{1/3},\,
\cr&
r_3 = R_2^{-2/3}R_3^{1/3}R_6^{-1/3}R_7^{1/3}R_8^{1/3},\,
r_4 = R_2^{1/3}R_3^{1/3}R_4 R_6^{2/3}R_7^{1/3}R_8^{1/3},\,
\cr&
r_5 = R_2^{1/3}R_3^{1/3}R_5 R_6^{2/3}R_7^{1/3}R_8^{1/3},\,
r_6 = R_2^{-2/3}R_3^{-2/3}R_6^{-1/3}R_7^{-2/3}R_8^{-2/3},\,
\cr&
r_7 = R_2^{1/3}R_3^{1/3}R_6^{-1/3}R_7^{1/3}R_8^{-2/3},\,
r_8 = R_2^{1/3}R_3^{1/3}R_6^{-1/3}R_7^{-2/3}R_8^{1/3},\,
\cr}}
If $R_2 = R_3 = \cdots = R_8 = 1$, then this transformation preserves
$\MT{7}$ and takes $R_1\rightarrow R_1^{-1}$.
In principle we should analyze this not for a right-angled $\MT{7}$
but for the $\MT{7}$ in, say, the example of appendix (C).
We believe that the conclusion will be the same, namely there
exists an extra element in $E_8(\BZ)$ which preserves $\MT{7}$
and takes $\rad\rightarrow 0$ to $\rad\rightarrow\infty$.

Although, mathematically, this $\Gamma'\subset E_8(\BZ)$ has
the required properties, it is not necessarily a duality in our case.
The reason is that we have to make a distinction between two
different ways of using $\ut\in E_7(\BZ)$ to compactify M-theory
down to 2+1D.
In case (i) we can mod out M-theory on $\MT{7}$ by $\ut$
already in 3+1D and then compactify the result to 2+1D.
In case (ii), which is our case, we compactify from 3+1D down
to 2+1D with a $\ut$-twist.
The subgroup $\Gamma'$ of $E_8(\BZ)$ which preserves the moduli of
$\MT{7}$ and the structure $\MT{7}\times \MS{1}$ is indeed a duality
in case (i). The disadvantage of case (i), however, is that we
cannot be sure the there are no new moduli from the twisted sectors.
In contrast,
case (ii) cannot be viewed as M-theory on $\MT{7}\times \MS{1}$
modded out by an element of $E_8(\BZ)$. Rather, what we really mod out
by is an element of $E_8(\BZ)$ compounded with a translation 
(by $1/6$ of the radius) along $\MS{1}$.
Not all the elements of $\Gamma'$ preserve the local operation of
translation along $\MS{1}$. Thus, we cannot be sure that $\Gamma'$ is
a duality.
In fact, if we were in any dimension higher than 2+1D we could most
likely rule out $\Gamma'$ as a duality group by studying 
the action on BPS states with momentum along $\MS{1}$.
In 2+1D $U(1)$ charges are confined and
there are no BPS states under local $U(1)$ symmetries 
and so the argument fails.

As an example, let us compare case (i) and case (ii) for a special
case in 5+1D. We can compactify M-theory on $\MT{4}$ and then further
down on $\MS{1}$ with a $\BZ_2$ twist.
If we first compactify (as in case (i))
on $\MT{4}/\BZ_2$ we can deform to a $K3$
and we get more moduli from the twisted sectors.
In case (ii) we do not get more moduli and the vector-multiplet
moduli space
is locally $SO(4,4)/(SO(4)\times SO(4))$, as we discussed in section (2.2)
(and see also \rDabHar).
Now we can ask what happens in the limit that the radius of $\MS{1}$
is kept fixed and the volume of $\MT{4}$ shrinks to zero.
In the first case we know from the duality between M-theory on $K3$
and heterotic on $\MT{3}$ that this becomes the weakly coupled
heterotic theory. Now, for an appropriate choice of Wilson line
along $\MS{1}$, the limit $\rad\rightarrow 0$ can be mapped
to the T-dual heterotic string in the limit $\rad\rightarrow\infty$
(replacing $E_8\times E_8$ with $SO(32)$).
This comes about because the M5-branes wrapped $N$ times
on the small $K3\times\MS{1}$ form light bound states which are then
interpreted as KK states of the large T-dual dimension.
So, although in this example $\rad\rightarrow 0$  is not dual
to another point with $\rad\rightarrow\infty$ on the same moduli space,
it is still a classical limit.
Now let us see what happens in case (ii).
It is likely that the M5-branes
wrapped $N$ times on the compact 5-manifold,
made by the $\MT{4}$ fibered over $\MS{1}$ with the twist, do not
form bound states.
To see why this is reasonable, let us first compactify the $(2,0)_N$
theory of the M5-branes on a small $\MT{4}$. To a good approximation,
we obtain 1+1D SYM with 16 supersymmetries. Now let us compactify this
on  $\MS{1}$ with a twist. The twist ``kills'' 4 scalars and so we are 
left with 0+1D SYM with $\SUSY{8}$ supersymmetries. This theory
does not have a bound state \refs{\rSetSte-\rKKN}.
Although the arguments for the bound state might not apply to this
case because some of the fields are 
compact, the conclusion is likely to be correct.\foot{I have
benefited from discussions with S. Sethi on this point.}




\subsec{Phase transitions}
Assuming that the moduli space is,
$$
\Gamma\backslash SU(2,1)/(SU(2)\times U(1)),
$$
we can explore the singularities of the moduli space
and the possible phase transitions that can occur when the moduli of
the supergravity theory reach these points.
The singular points are fixed points of elements of $\Gamma$.
An example of such a fixed point is furnished by the element $S$ of
\radinv\ that inverts $\Phi +2ie \rad$ at $\sc_1 = \sc_2 = 0$.
From \gact\ it can easily be checked that the only fixed
point in the region \restzz\  is at $(z_1,z_2) = (1,0)$.
This corresponds to $\rad = {1\over 2}$
and $\Phi = \sc_1 = \sc_2 = 0$. 
It follows  that the local structure near
that point is $\MR{4}/\BZ_2$. This is the same structure
as 2+1D $\SUSY{4}$ QCD with $N_f=2$ quarks \refs{\rSeiIRD,\rSWGDC}.
The latter theory has another phase emanating from the singularity.
Close to the singularity the structure of the other phase
looks like $\MR{4}/\BZ_2$ \rSeiIRD.
 It has 3 compact parameters which parameterize $SO(3)$
and 1 non-compact parameter $0 < \rho <\infty$.
We expect the same structure for small $\rho$ in our case as well.
At the singular point $\rad = {1\over 2}$ and $\Phi = \sc_1 = \sc_2$
the low-energy physics is described by a conformal theory whose
moduli space is $\MR{4}/\BZ_2$. At low-energies the coupling
to gravity can be ignored and since QCD with $N_f = 2$ quarks (or
its mirror \rIntSei) is
the only CFT which we know to possess this kind of singularity, it is
natural to suspect that our supergravity theory
is described at low-energies, at this particular singular point,
by the same theory.

 We therefore conclude that there is another phase 
of supergravity emanating from that
point. At the other phase the original supergravity variables
$\rad,\Phi,\sc_1,\sc_2$ are massive but instead we get 4 new moduli.
One of them is $\rho$.
We only know the structure of the moduli space of the other
phase near $\rho = 0$.
 For $\rho$ of the order of $M_p^{1/2}$,
gravity mixes with the CFT again.
We know from \rNTW\ that the moduli space has to be quaternionic.
It is plausible that 
in the limit $\rho\rightarrow\infty$ (assuming the
moduli-space of the other phase is non-compact)
the description of the low-energy
modes of the theory becomes classical again,  but with a totally
different description. 



\newsec{Discussion}
In the first part of this paper we have studied compactifications
of M-theory with U-duality twists. We have seen that many moduli spaces
from the list of \rNTW\ can be realized in this way.
We have then proceeded to study a particular realization of $\SUSY{4}$ 2+1D
supergravity.
We have given arguments for the existence of a particular
twist with an isolated fixed point and which preserves ${1\over 4}$
of the supersymmetry.
We have given the explicit form of the twist $\ut$ in 
$SU(8)\subset E_7(\BR)$ and we have shown that its characteristic
polynomial has integral coefficients.
In appendix (C) we will construct it more explicitly.

We conjectured that the moduli space of this theory
is given by $\Gamma'\backslash SU(2,1)/U(2)$ where 
the discrete U-duality group $\Gamma'$ is an extension
of the classical discrete group $\Gamma_{cl}$ by a particular S-duality 
element. The quantum moduli space has singular points.
We conjectured that the low-energy modes at the singular point in moduli 
space are described by the only known 2+1D
conformal field-theories \rIntSei\ with  these types of singularity.
We concluded that other phases emanate from these points.
These conjectures rest on the assumption that quantum corrections
do not modify the local structure of the moduli space.


In case the new phase does exist,
it is an open problem whether the moduli space
of the new phase is compact or not.
If it is non-compact then far away from the singularity
it might be described 
by a ``classical limit'' of some sort.
Perhaps the theory grows more dimensions,
perhaps it becomes a weakly coupled string theory, or perhaps it 
becomes a completely new classical limit
which we have never encountered before.

Looking farther ahead, it would be interesting to understand
``where'' the new phases of gravity discussed in this paper 
``sit'' with respect to the more exotic
phases of gravity. These are the topological phase described in
\rWitTOP\ and the phase with $\ev{g_{\u\v}} = 0$ suggested in
\rPolyak.


\bigbreak\bigskip\bigskip
\centerline{\bf Acknowledgments}\nobreak
I am very grateful  to A.M. Polyakov, S. Sethi, E. Witten
and especially N. Seiberg
for discussions and to A. Strominger for correspondence.


\appendix{A}{Quick review of 2+1D moduli spaces}
The results of \rNTW, for more than $\SUSY{4}$
supersymmetries, were summarized in the following table
(extracted from table (3) of \rNTW).
\bigskip
\vbox{\settabs 3 \columns
\+${\cal N}$  & Dim   &   $G/K$              \cr
\+ 16         & $128$   &  $E_{8(+8)}/SO(16)$  \cr
\+ 12         & $64$    &  $E_{7(-14)}/(SO(12)\otimes SO(3))$ \cr
\+ 10         & $32$    &  $E_{6(-12)}/(SO(10)\otimes SO(2))$ \cr
\+ 9          & $16$    &  $F_{4(-20)}/SO(9)$ \cr
\+ 8          & $8k$    &  $SO(8,k)/(SO(8)\otimes SO(k))$ \cr
\+ 6          & $8k$    &  $SU(8,k)/S(U(4)\otimes U(k))$ \cr
\+ 5          & $8k$    &  $Sp(2,k)/(Sp(2)\otimes Sp(k))$ \cr
}

Here ${\cal N}$ is the number of supersymmetries, ``Dim'' is the dimension
of the moduli space and $G/K$ is the local form of the moduli space.

 For the case of $\SUSY{4}$ the moduli space has to be quaternionic.
 For a 4-dimensional manifold this means that the curvature satisfies,
$$
0 = R_{ijkl} + {1\over 2}\sqrt{g}{R_{ij}}^{mn}\epsilon_{mnkl}
 + g_{ik}g_{jl}-g_{il}g_{kl} + \sqrt{g}\epsilon_{mnkl}.
$$


\appendix{B}{The fixed point of the U-duality element}
In section (3), we needed a U-duality element
$\ut\in E_7(\BZ)$ and an element $g\in E_{7(7)}(\BR)$ and
$\ko\in SU(8)$ such that,
$$
g \ko g^{-1} = \ut,
$$
and the eigenvalues of $\ko$ are,
\eqn\spko{
(\underbrace{e^{{{i\pi}\over 3}},\dots,
   e^{{{i\pi}\over 3}}}_{6\ {\rm times}}, 1,1).
}
In this section we will make some observations on such a $\ut$,
independent of a particular realization.
We will start by finding the charge in M-theory on $\MT{7}$ that
is fixed by $\ut$.
In particular, we wish to know how much SUSY the corresponding charged
particle preserves.

We start with preliminaries.

\subsec{Central charge formula for M-theory on $\MT{7}$}
The central charge formula for M-theory on $\MT{7}$ has been explained
in \rWitVAR.
The SUSY charges are in the representation,
$$
(\rep{2},\rep{8}) + (\rep{\mybar{2}},\rep{\mybar{8}}),
$$
of $SO(3,1)\times SU(8)$.
The charge is a vector $q$
in the representation $\rep{56}$ of $E_{7}(\BR)$.
The central charge $Z$ is a vector in the $\rep{28}+\rep{\mybar{28}}$
 of $SU(8)\subset E_7(\BR)$.
We pick a map $T$ from $\rep{56}$ of $E_7$ to 
$\rep{28}+\rep{\mybar{28}}$ of $SU(8)$. The map is required to satisfy,
$$
T\circ \Omega = \Omega\circ T,\qquad \Omega\in SU(8)\subset E_7(\BR)
$$
Let $g$ be a representative of the $E_7(\BR)/SU(8)$ point corresponding
to the moduli. $g$ is defined up to $g\rightarrow g\Omega$ with
$\Omega\in SU(8)$.
The relation between the central charge and the charge vector is,
$$
\wZ = T g^{-1} q.
$$

\subsec{Number of unbroken supersymmetries for BPS states}
Now suppose a state has a central charge matrix of $\wZ$.
What is the maximal number of supersymmetries that it can preserve?
$\wZ$ can be thought of as an $8\times 8$ antisymmetric matrix.
The representation $\rep{28}+\rep{\mybar{28}}$ is made
from $Z$ and $Z^{*}$.
So now we can write the commutation relations,
\eqn\comrelqq{\eqalign{
\acom{Q_{i\a}}{Q_{\mybar{j}\dot{\b}}} &=
\delta_{i\mybar{j}}\sigma^\mu_{\a\dot{\b}} P_\u,  \cr
\acom{Q_{i\a}}{Q_{j\b}} &=
Z_{ij}\epsilon_{\a\b},\cr
\acom{Q_{\mybar{i}\dot{\a}}}{Q_{\mybar{j}\dot{\b}}} &=
\mybar{Z}_{\mybar{i}\mybar{j}}\epsilon_{\dot{\a}\dot{\b}}.\cr
}}
As a Matrix this is,
$$
\pmatrix{
0 & 0 & E\delta_{\mybar{i}j} & \mybar{Z}_{\mybar{i}\mybar{j}} \cr
0 & 0 & -{Z}_{{i}{j}} & E\delta_{i\mybar{j}} \cr
E\delta_{i\mybar{j}} & {Z}_{{i}{j}} & 0 & 0\cr
-\mybar{Z}_{\mybar{i}\mybar{j}} & E\delta_{\mybar{i}j} & 0 & 0\cr
}.
$$
The columns and rows
are in the order $\dot{1}\mybar{j},2j, 1j, \dot{2}\mybar{j}$
We define the $16\times 16$ Hermitian matrix,
$$
A \equiv\pmatrix{
E\delta_{\mybar{i}j} & \mybar{Z}_{\mybar{i}\mybar{j}} \cr
-{Z}_{{i}{j}} & E\delta_{i\mybar{j}} \cr}
= \pmatrix{
E\,\Id & -Z^\dagger \cr -Z & E\, \Id\cr}.
$$
and also,
$$
B = \pmatrix{
E\,\Id & Z \cr Z^\dagger & E\, \Id \cr}.
$$
The $32\times 32$ matrix becomes,
$$
\pmatrix{0 & A \cr B & 0 \cr}.
$$
We need to know how many zeroes $A$ and $B$ have together.
Since $A$ is the complex conjugate of $B$ it is enough to count the
number of zeroes of $A$.
Thus, we are looking for,
$$
0 = \pmatrix{
E\,\Id & -Z^\dagger \cr -Z & E\, \Id\cr}
\pmatrix{u \cr v} \Longrightarrow (Z^\dagger Z - E^2\Id) u = 0.
$$
It can easily be checked that the number of supersymmetries
that are preserved by the state is twice the dimension
of the eigen-space of the largest  eigenvalue of $Z^\dagger Z$.
What do we know about $Z$?
Let $g$ be the fixed point of the U-duality twist $\ut$.
Thus,
$$
\ut g = g\ko.
$$
Let $q$ be a vector of charges which is invariant under $\ut$,
$$
\ut q = q.
$$
It follows that,
$$
\wZ = T g^{-1} q = T g^{-1}\ut^{-1} q = T \ko^{-1} g^{-1} q
 = \ko^{-1} T g^{-1} q = \ko^{-1} \wZ.
$$
This is in the representation $\rep{28} + \rep{\mybar{28}}$.
To bring this back to the $8\times 8$ antisymmetric $Z_{ij}$
we have,
$$
Z = \ko Z \ko^t.
$$
Taking $\ko$ as given in \spko\ we see that $Z$ is,
$$
Z = \pmatrix{
0 & 0 & 0 & 0 & 0 & 0 & 0 & 0 \cr
0 & 0 & 0 & 0 & 0 & 0 & 0 & 0 \cr
0 & 0 & 0 & 0 & 0 & 0 & 0 & 0 \cr
0 & 0 & 0 & 0 & 0 & 0 & 0 & 0 \cr
0 & 0 & 0 & 0 & 0 & 0 & 0 & 0 \cr
0 & 0 & 0 & 0 & 0 & 0 & 0 & 0 \cr
0 & 0 & 0 & 0 & 0 & 0 & 0 & Z_0 \cr
0 & 0 & 0 & 0 & 0 & 0 & -Z_0 & 0 \cr}.
$$
Thus, $Z Z^\dagger$ has two eigenvalues of $|Z_0|^2$ and 6 eigenvalues
of zero.
We see that a BPS state with the corresponding fixed charge
preserves ${1\over 8}$ of the supersymmetry.
Out of the broken supersymmetry generators, $24 = 6\times 4$ 
transform under $\ut$ with a phase and $4$ generators are invariant.





\appendix{C}{Explicit construction of $\ut$}
In section (3.2) we needed
an element $\ut\in E_7(\BZ)$ and an element $g\in E_{7(7)}(\BR)$ such that
$g\circ \ut \circ g^{-1}$ is equal to $\ko\in SU(8)$ in \koeig.

We will now give one explicit construction for such an element 
$\ut\in E_7(\BZ)$.
For the construction it is convenient to view the compactification
as type-II on $\MT{6}$.
Let us set all the RR-fluxes to zero. The moduli space
is,
$$
\left(SO(6,6,\BZ)\backslash SO(6,6,\BR) / (SO(6)\times SO(6))\right)
\otimes
\left(SL(2,\BZ)\backslash SL(2,\BR)/SO(2)\right).
$$
The first factor corresponds to the metric and NS-NS 2-form
fluxes on $\MT{6}$
while the second part corresponds to the parameter,
$$
\chi \equiv {{i V}\over {\lam_s^2}} + \widetilde{C},
$$
where $\widetilde{C}$ is the NS-NS 6-form flux (dual of the NS-NS 2-form)
on $\MT{6}$.
$V$ is the overall volume of $\MT{6}$ and $\lam_s$ is the string
coupling constant.
We will denote the $SO(2)\subset SL(2,\BR)$ by
$U(1)_{\chi}$.
The subgroup of the 
U-duality group that preserves the condition that the RR-fluxes
are zero is,
$$
SL(2,\BZ)\times SO(6,6,\BZ).
$$
At special points in the moduli space, a finite subgroup of
the T-duality group $SO(6,6,\BZ)$ becomes a symmetry. The finite
subgroup can then be identified with a discrete subgroup of
a cover of $SO(6)\times SO(6)$. This cover is $SU(4)\times SU(4)$.
A symmetry element which can be realized as a geometrical
transformation of $\MT{6}$ can be embedded in the diagonal $SO(6)$.
Similarly, at special points in the moduli space, a finite subgroup
of the S-duality group $SL(2,\BZ)$ becomes a symmetry. It can then
be identified with a discrete subgroup of $U(1)_\chi$.
Now we can embed $SU(4)\times SU(4)\times U(1)_\chi$ inside $SU(8)$
which we have used to write down \koeig.
Under this embedding,
$$
\rep{8} = \rep{4}_{+} + \rep{4}_{-}.
$$
Now let us take a special $\MT{6}$ of the form
$\MT{2}_a\times \MT{2}_b\times\MT{2}_c$.
To make the discussion clearer, we have added subscripts to
the $\MT{2}$'s.
Let us denote the complex structure of each $\MT{2}$ by $\tau$ (with
an appropriate subscript) and the combination 
${B\over {2\pi}} + i A$ by $\rho$.
Here $B$ is the NS-NS 2-form and $A$ is the area.
Type-II on $\MT{2}$ has a moduli space of,
$$
(SL(2,\BZ)\backslash SL(2,\BR) / U(1)_\rho)\times 
(SL(2,\BZ)\backslash SL(2,\BR) / U(1)_\tau)
$$
corresponding to the pair $(\rho,\tau)$.
If we write the moduli space in the form,
$$
SO(2,2,\BZ)\backslash SO(2,2,\BR) / (SO(2)_1\times SO(2)_2),
$$
then $U(1)_\tau$ is the diagonal combination of $SO(2)_1$ and $SO(2)_2$
while $U(1)_\rho$ is the combination with $SO(2)_2$ inverted.
As before, at special points in the moduli space we can identify
the finite subgroup of the U-duality group which preserves the point
and the structure $\MT{2}_c\times\MT{2}_b\times\MT{2}_c$
with a  discrete subgroup of,
\eqn\allus{
U(1)_{\rho_a}\times U(1)_{\tau_a}\times
U(1)_{\rho_b}\times U(1)_{\tau_b}\times
U(1)_{\rho_c}\times U(1)_{\tau_c}\times
U(1)_\lam \subset SU(8).
}
The representation $\rep{8}$ decomposes as,
\eqn\uuudec{\eqalign{
(+++++++) \oplus 
(++----+) \oplus
&
(--++--+) \oplus 
(----+++) \oplus 
\cr
(+-+-+--) \oplus 
(+--+-+-) \oplus
&
(-++--+-) \oplus 
(-+-++--)
\cr
}}
Now we take $\MT{2}_a$ to have $\tau_a =\rho_a = e^{-{{2 \pi i}\over 3}}$
and $\MT{2}_b$ and $\MT{2}_c$ to have,
$$
\tau_b = \tau_c = \rho_b = \rho_c = i.
$$
We also take,
$$
\chi = e^{-{{2 \pi i}\over 3}}.
$$
The U-duality element $\ut$ is a combination of,
\eqn\utforx{\eqalign{
\tau_b\rightarrow -{1\over \tau_b},\,\,
\rho_b\rightarrow -{1\over \rho_b},&\qquad
\tau_c\rightarrow -{1\over \tau_c},\,\,
\rho_c\rightarrow -{1\over \rho_c},\cr
\tau_a\rightarrow -1-{1\over \tau_a},\,\,
\rho_a\rightarrow -1-{1\over \rho_a},&\cr
}}
and we top it with an S-duality transformation,
$$
\chi\rightarrow -1 - {1\over \chi}.
$$
This corresponds to the following elements in \allus:
$$
(e^{{{-\pi i}\over 3}}, e^{{{-\pi i}\over 3}},
e^{{{\pi i}\over 4}},
e^{{{\pi i}\over 4}},
e^{{{\pi i}\over 4}},
e^{{{\pi i}\over 4}},
 e^{{{-\pi i}\over 3}}).
$$
Note that since all of these transformations can be U-conjugated
to rotations of a $\MT{2}$ and since the $U(1)$ charges are
measured with respect to the fermions, the phases are half
the rotation angle. (The overall $(-)^F$ is a symmetry of M-theory.)
Using \uuudec\ we calculate the eigenvalues
of $\ko$ in $\rep{8}$ of $SU(8)$ to be,
$$
(1,1,
e^{{{\pi i}\over 3}}, e^{{{\pi i}\over 3}},
e^{{{\pi i}\over 3}}, e^{{{\pi i}\over 3}},
e^{{{\pi i}\over 3}}, e^{{{\pi i}\over 3}}).
$$

\subsec{The invariant charge}
We can now calculate the invariant charge corresponding to the 
${1\over 8}$-BPS particle (see appendix B).
The respresentation of the charges is the $\rep{56}$ of $E_7(\BZ)$.
Under,
$$
SO(6,6,\BZ)\times SL(2,\BZ),
$$
it decomposes as,
$$
\rep{56} = (\rep{12},\rep{2}) + (\rep{32},\rep{1}),
$$
(see \rWitVAR).
Under $SO(2,2)_a\times SO(2,2)_b\times SO(2,2)_c\times SL(2)_\chi$
it decomposes as,
\eqn\fsdec{\eqalign{
\rep{56} =&
(\rep{4},\rep{1},\rep{1},\rep{2}) +
(\rep{1},\rep{4},\rep{1},\rep{2}) +
(\rep{1},\rep{1},\rep{4},\rep{2}) +
\cr &
(\rep{2},\rep{2},\rep{2},\rep{1}) +
(\rep{2},\rep{2}',\rep{2}',\rep{1}) +
(\rep{2}',\rep{2},\rep{2}',\rep{1}) +
(\rep{2}',\rep{2}',\rep{2},\rep{1})
}}
Under $U(1)_{\rho_a}\times U(1)_{\tau_a}$ of \allus,
$\rep{4},\rep{2}$ and $\rep{2}'$ of $SO(2,2)_a$ decompose as,
\eqn\sorepde{\eqalign{
\rep{4} =& (+1,+1)\oplus(+1,-1)\oplus(-1,+1)\oplus (-1,-1),\cr
\rep{2} =& (+1,0)\oplus(-1,0),\cr
\rep{2}' =& (0,+1)\oplus(0,-1),\cr
}}
(If we replace type-IIA with type-IIB, we have to replace $\rho_a$
with $\tau_a$.)
We see that two vectors out of,
$(\rep{4},\rep{1},\rep{1},\rep{2})$,
are invariant under $\ut$.
The corresponding BPS states are combinations of strings wrapped
on $\MT{2}_a$ and KK states along the sides of $\MT{2}_a$
together with their duals which are NS5-branes and KK monopoles.


\appendix{D}{The $SU(2,1)/(SU(2)\times U(1))$ moduli space}
We have seen that the target-space  with the metric,
\eqn\appmet{
ds^2 = {{1}\over {e^4\rad^4}} (d\sc_1^2 + 2\Phi d\sc_1 d\sc_2 
   + (e^2\rad^2 + \Phi^2) d\sc_2^2) + 4{{d\rad^2}\over {\rad^2}}
 + {1\over {e^2\rad^2}}d\Phi^2,
}
is equivalent to the space $SU(2,1)/(SU(2)\times U(1))$ and the latter can
be described in terms two complex coordinates $(z_1,z_2)$,
with the action of an $SU(2,1)$ matrix,
$$
g = \pmatrix{
g_{11} & g_{12} & g_{13} \cr
g_{21} & g_{22} & g_{23} \cr
g_{31} & g_{32} & g_{33} \cr},
$$
given by,
$$
g : (z_1, z_2)\rightarrow
\left(
{{g_{11} z_1 + g_{12} z_2 + g_{13}}
       \over {g_{31} z_1 + g_{32} z_2 + g_{33}}},
{{g_{21} z_1 + g_{22} z_2 + g_{23}}
       \over {g_{31} z_1 + g_{32} z_2 + g_{33}}}\right).
$$
The purpose of this appendix is to describe the precise mapping.

\subsec{Killing vectors}
The metric \appmet\ possesses a group of isometries 
which  can be described by the following infinitesimal
transformation laws,
\eqn\klvec{\eqalign{
\delta \sc_1 &=
\epsilon_1\{{3\over 2}\Phi^2 \sc_2-{{1}\over 2}\sc_2^3\}
+ \epsilon_2\{
  {{3}\over{8}}\Phi^4
+ 2 e^2\rad^2 \Phi^2
+ 2e^4\rad^4
- 2 \sc_1^2
+ {{3}\over{4}}\Phi^2 \sc_2^2
- {{1}\over{8}}\sc_2^4
\}
\cr &
- 2 \epsilon_3 \sc_1
- \epsilon_4\{
  \Phi^3
+ 2e^2\rad^2 \Phi
+ 2\sc_1 \sc_2
\}
- \epsilon_5\sc_2
+ \epsilon_6\{
  {1\over {2}}\Phi^2 - {{1}\over 2}\sc_2^2\}
+ \epsilon_8,
\cr
\delta \sc_2 &=
-\epsilon_1\{
  2\sc_1 + 3\Phi \sc_2\}
-\epsilon_2\{
  {1\over {2}}\Phi^3
+ 2e^2 \rad^2\Phi
+ 2 \sc_1 \sc_2
+ {{3}\over{2}} \Phi \sc_2^2
\}
\cr &
- \epsilon_3 \sc_2
+\epsilon_4\{
  {3\over {2}} \Phi^2
+ 2 e^2\rad^2
- {{1}\over{2}}\sc_2^2
\}
 - \epsilon_6\Phi
 + \epsilon_7
\cr
\delta \Phi &=
\epsilon_1\{
  2e^2\rad^2
-{{1}\over{2}} \Phi^2
+ {{3}\over{2}} \sc_2^2
\}
+\epsilon_2\{
  2 e^2\rad^2 \sc_2
- {{1}\over{2}} \Phi^2 \sc_2
- 2 \Phi \sc_1
+ {{1}\over{2}} \sc_2^3
\}
\cr &
- \epsilon_3\Phi
+\epsilon_4\{
 2\sc_1 - \Phi\sc_2\}
+ \epsilon_5 + \epsilon_6\sc_2,
 \cr
\delta \rad &=
-\epsilon_1 \Phi \rad
-\epsilon_2\{
 2\rad \sc_1 + \Phi \rad \sc_2\}
- \epsilon_3\rad 
- \epsilon_4\rad\sc_2.
\cr
}}
Here $\epsilon_1,\dots,\epsilon_8$ are arbitrary.

Writing,
$$
\delta \equiv \sum_1^8\epsilon_i\delta_i,
$$
we can calculate the commutation relations,
\eqn\comdelx{\eqalign{
[ \delta_1, \delta_2 ] = 0, \qquad
[ \delta_1, \delta_3 ] = -\delta_1,&\qquad
[ \delta_1, \delta_4 ] = 2\delta_2, \qquad
[ \delta_1, \delta_5 ] = \delta_3,\cr
[ \delta_1, \delta_6 ] = \delta_4, \qquad
[ \delta_1, \delta_7 ] = 3\delta_6,&\qquad
[ \delta_1, \delta_8 ] = -2\delta_7, \qquad
[ \delta_2, \delta_3 ] = -2\delta_2,\cr
[ \delta_2, \delta_4 ] = 0, \qquad
[ \delta_2, \delta_5 ] = -\delta_4, &\qquad
[ \delta_2, \delta_6 ] = 0, \qquad
[ \delta_2, \delta_7 ] = \delta_1,\cr
[ \delta_2, \delta_8 ] = 2\delta_3, \qquad
[ \delta_3, \delta_4 ] = \delta_4, &\qquad
[ \delta_3, \delta_5 ] = -\delta_5, \qquad
[ \delta_3, \delta_6 ] = 0,\cr
}}
\eqn\comdely{\eqalign{
[ \delta_3, \delta_7 ] = -\delta_7,\qquad
[ \delta_3, \delta_8 ] = -2\delta_8,&\qquad
[ \delta_4, \delta_5 ] = -3\delta_6, \qquad
[ \delta_4, \delta_6 ] = -\delta_1,\cr
[ \delta_4, \delta_7 ] = \delta_3,\qquad
[ \delta_4, \delta_8 ] = 2\delta_5,&\qquad
[ \delta_5, \delta_6 ] = \delta_7, \qquad
[ \delta_5, \delta_7 ] = -\delta_8,\cr
[ \delta_5, \delta_8 ] = 0,\qquad
[ \delta_6, \delta_7 ] = \delta_5,&\qquad
[ \delta_6, \delta_8 ] = 0, \qquad
[ \delta_7, \delta_8 ] = 0,\cr
}}
These generators form the Lie algebra of $SU(2,1)$.
We can represent the algebra as the set of $3\times 3$ matrices $A$
which satisfy,
$$
A^\dagger J = -J A,
$$
where,
$$
J = \pmatrix{
0  & 0 & -1 \cr
0  & 2 &  0 \cr
-1 & 0 &  0 \cr}.
$$
The generators can be represented as,
\eqn\genrepx{\eqalign{
\delta_1 =
\pmatrix{
           0 & 0 & 0 \cr
-{1\over 2}i & 0 & 0 \cr
           0 & i & 0 \cr},&\qquad
\delta_2 =
\pmatrix{
            0 & 0 & 0 \cr
            0 & 0 & 0 \cr
 -{1\over 2}i & 0 & 0 \cr},\cr
\delta_3 =
\pmatrix{
-1 & 0 & 0 \cr
 0 & 0 & 0 \cr
 0 & 0 & 1 \cr}, &\qquad
\delta_4 =
\pmatrix{
          0 & 0 & 0 \cr
-{1\over 2} & 0 & 0 \cr
          0 &-1 & 0 \cr},\cr
}}
\eqn\genrepy{\eqalign{
\delta_5 =
\pmatrix{
 0 & 2i& 0 \cr
 0 & 0 &-i \cr
 0 & 0 & 0 \cr},&\qquad
\delta_6 =
\pmatrix{
-{i\over 3} &       0       & 0 \cr
          0 & {{2i}\over 3} & 0 \cr
 0          &       0       &-{i\over 3} \cr},\cr
\delta_7 =
\pmatrix{
 0 &-2 & 0 \cr
 0 & 0 &-1 \cr
 0 & 0 & 0 \cr},&\qquad
\delta_8 =
\pmatrix{
 0 & 0 & 4i \cr
 0 & 0 &  0 \cr
 0 & 0 &  0 \cr}.\cr
}}

\subsec{Integrated forms}
We will now write down some group element actions of the
form $e^{t \delta_i}$.
It is easy to see that,
\eqn\integfa{\eqalign{
e^{t\delta_8}(\sc_1,\sc_2,\Phi,\rad) &=
   (\sc_1 + t,\sc_2, \Phi,\rad),\cr
e^{t\delta_7}(\sc_1,\sc_2,\Phi,\rad) &=
   (\sc_1,\sc_2 + t, \Phi,\rad),\cr
e^{t\delta_5}(\sc_1,\sc_2,\Phi,\rad) &=
   (\sc_1 - t\sc_2,\sc_2, \Phi + t,\rad),\cr
e^{t\delta_3}(\sc_1,\sc_2,\Phi,\rad) &=
   (e^{-2t}\sc_1,e^{-t}\sc_2, e^{-t}\Phi,e^{-t}\rad).\cr
}}
It is also easy to integrate,
\eqn\integfb{\eqalign{
e^{t\delta_6}\sc_1 &= \sc_1
    +{1\over 4}(\Phi^2 - \sc_2^2)\sin 2t
       +{1\over 2}\Phi\sc_2 (1-\cos 2t),\cr
e^{t\delta_6}\sc_2 &= \sc_2 \cos t - \Phi\sin t,\cr
e^{t\delta_6}\Phi &= \Phi \cos t + \sc_2\sin t,\cr
e^{t\delta_6}\rad &= \rad.\cr
}}

Now let us check the flow generated by $\delta_1 - {1\over 2}C^2\delta_5$.
\eqn\flowfiv{\eqalign{
{{d\sc_1}\over {dt}} &= 
   {3\over 2}\Phi^2\sc_2 - {{1}\over 2}\sc_2^3 
          + {1\over 2}C^2\sc_2,\cr
{{d\sc_2}\over {dt}} &= 
  -2\sc_1 -3\Phi\sc_2,\cr
{{d\Phi}\over {dt}} &=
  2 e^2\rad^2 - {1\over 2}\Phi^2 + {{3}\over 2}\sc_2^2 
         - {1\over 2}C^2,\cr
{{d\rad}\over {dt}} &=
          -\Phi\rad.\cr
}}
Let us solve it for $\sc_1=\sc_2=0$.
We then define $W = \Phi + 2 i e \rad$. The equations are,
$$
{{dW}\over {dt}} = -{1\over 2}W^2 -{1\over 2}C^2.
$$
Thus,
$$
W = 
   {{W_0 - C\tan {1\over 2}C t} \over 
      {1 + {{W_0}\over {C}}\tan {1\over 2}C t}}.
$$
For $C t = \pi$ this becomes the
  transformation $W_0\rightarrow -C^2/W_0$ which we will interpret
as a strong/weak duality.
For $C=1$ The matrix representation gives,
$$
e^{\pi(\delta_1-{1\over 2}\delta_5)} = \pmatrix{
0 & 0 &  1\cr
0 &-1 &  0\cr
1 & 0 &  0\cr}.
$$

\subsec{Equations for the change of variables}
We can now write down the equations for the change of variables
from the $(\sc_1,\sc_2,\Phi,\rad)$ variables to $(z_1,z_2)$.
To do this we will use a combination of $\delta_3$ and
$\delta_5,\delta_7,\delta_8$ to move from one point on the manifold
to another,
\eqn\movdel{
e^{t\delta_3} e^{b\delta_5}
e^{a_1\delta_8 + a_2\delta_7}
(\sc_1,\sc_2,\Phi,\rad)
=
(e^{-2t}\sc_1 -  b e^{-t}\sc_2 + {a_1},
 e^{-t}\sc_2 + a_2, e^{-t}\Phi + b, e^{-t}\rad).
}
We now take,
\eqn\nowtak{\eqalign{
t &= \log {{\rad}\over {\wrad}},\cr
b &= \wPhi - {{\rad}\over {\wrad}}\Phi,\cr
a_2 &= \wsc_2 - {{\rad}\over {\wrad}}\sc_2,\cr
a_1 &= \wsc_1 - {{\rad^2}\over {\wrad^2}}\sc_1 + 
       {{\rad}\over {\wrad}}b \sc_2.\cr
}}
The rescaling transformation $e^{t\delta_3}$ also makes it
obvious that the classical regime $\rad\rightarrow\infty$
is given by
$$
\Re z_1 - |z_2|^2 \rightarrow \infty.
$$
Note that neither of $\delta_5,\delta_7,\delta_8$ change
the value of $\Re z_1 - |z_2|^2$, as it should be.


\def\bmu{\bar{\mu}}

\appendix{E}{On the structure of $\Gamma$ for $\tau= 2i$}
We have defined $\Gamma'\subset SU(2,1)$ as the subgroup generated by,
\eqn\gensh{
g_2 = \pmatrix{
1 & 2 &  1  \cr 
0 & 1 &  1  \cr
0 & 0 &  1  \cr},\qquad
g_3 = \pmatrix{
1 &2\mybar{\tau}&  |\tau|^2  \cr 
0 & 1 &  \tau  \cr
0 & 0 &  1  \cr},
}
and,
$$
S = \pmatrix{
0 &  0 &  1 \cr
0 & -1 &  0 \cr
1 &  0 &  0 \cr}.
$$
In this appendix we will prove that, for the special
case of $\tau=2i$, adding $S$ does not cause
any unwanted identifications in the classical region.
I do not know how to generalize the proof to other $\tau$'s.
Perhaps, embedding $\Gamma'$ in $E_8(\BZ)$ as the subgroup that
preserves $\ut$ would be a good tactic.
Translated to matrices, this statement means that there
is no element $g\in\Gamma - \Gamma_{cl}$ such that 
the vector,
$$
\pmatrix{1 \cr 0 \cr 0 \cr}
$$
is an eigen-vector of $g$.

\subsec{The boundary of $SU(2,1)/(SU(2)\times U(1))$}
To understand  $\Gamma$, it is helpful to analyze its action on the
boundary of $SU(2,1)/(SU(2)\times U(1))$.
This boundary is given by
$\Re {z_1} = |z_2|^2$ (equation \restzz).
Adding a point at infinity, the boundary
is easily seen to be $\MS{3}$ (this is easier to see in
a diagonal metric of signature $(1,1,-1)$).

The classical region is given by $\Re z_1 \gg |z_2|^2$ but it is
easy to see that this is just the point at infinity that was added.
Thus, in order to prove that $\Gamma$ does not make extra identifications
in the classical region, we need to analyze the action of $\Gamma$
on the boundary.

\subsec{The structure of $\Gamma$}
We set $\tau=2i$ and find,
$$
g_1 = g_3^{-1}  g_2^{-1}  g_3 g_2 =
\pmatrix{
1 & 0 & 8i \cr
0 & 1 &  0  \cr
0 & 0 &  1  \cr}.
$$
Now let us introduce,
$$
g(k,l,m) \equiv
g_1^m g_2^k g_3^l  = \pmatrix{
1 & 2 (k - 2il) & (k^2 + 4l^2) + 4 i  (kl + 2m) \cr
0 &     1      & (k + 2i l)                   \cr
0 &     0      &      1                      \cr}.
$$
We have the rule,
$$
g(k_1, l_1, m_1) g(k_2, l_2, m_2) 
= g(k_1 + k_2, l_1 + l_2, m_1 + m_2 + k_2 l_1).
$$

A generic term in $\Gamma$ is of one of the three forms,
\eqn\genericg{\eqalign{
h_1 &=
S g(k_1, l_1, m_1) S g(k_2, l_2, m_2) S g(k_3, l_3, m_3) \cdots
S g(k_r, l_r, m_r),\cr
h_2 &=
 g(k_1, l_1, m_1) S g(k_2, l_2, m_2) S g(k_3, l_3, m_3) \cdots
S g(k_r, l_r, m_r),\cr
h_3 &=
 g(k_1, l_1, m_1) S g(k_2, l_2, m_2) S g(k_3, l_3, m_3) \cdots
S g(k_r, l_r, m_r) S.\cr
h_4 &=
 S g(k_1, l_1, m_1) S g(k_2, l_2, m_2) S g(k_3, l_3, m_3) \cdots
S g(k_r, l_r, m_r) S.\cr
}}
We can also require that no
$g(0,\pm 1,0)$ ever appear between two $S$'s because we could then
reduce the number of $S$'s (denoted by $r$) according to,
$$
S g(\pm 1,0,0) S = g(\mp 1, 0, 0) S g(\mp 1, 0, 0).
$$
Now we need a lemma.
\vskip 0.2cm
{\bf Lemma:}
\vskip 0.2cm
Let $g$ be given by,
\eqn\sgsg{
g = 
S g(k_1, l_1, m_1) S g(k_2, l_2, m_2) S g(k_3, l_3, m_3) \cdots
S g(k_r, l_r, m_r),
}
Such that for all $j$,
$$
(k_j,l_j,m_j)\neq (\pm 1,0,0),
$$
Let,
$$
x = (z_1 = |z_2|^2 + i w, z_2),\qquad z_2\in\BC,\qquad w\in\BR,
$$
be  a point on the boundary that satisfies
\eqn\zbnd{
0 < |z_2|< 1,\qquad
-{1\over 2} < w < {1\over 2},
}
Let $x'$ be the result of the action of $g$ on $x$ (which is
still a point on the boundary),
$$
x' \equiv g x = (|z_2'|^2 + i w', z_2').
$$
then $0 < |z_2'| < 1$ and $|w'| < {1\over 2}$.
\vskip 0.5cm
{\bf Proof:}
\vskip 0.5cm
It is sufficient to check for,
$$
g = S g(k,l,m).
$$
Let us denote,
$$
\mu = k + 2 i l,\qquad \zeta = 4 k l + 8m.
$$
 From \gact\ we find,
\eqn\newzz{\eqalign{
z_2' &=
-{{\mu + z_2}\over {|z_2|^2 + 2\bmu z_2 + |\mu|^2 + i (w +\zeta)}},\cr
 w' &=
{{w + \zeta + 2 \Im (\bmu z_2)}\over
  {\left| |z_2|^2 + 2\bmu z_2 + |\mu|^2 + i (w +\zeta)\right|^2}},\cr
}}
First, let us assume that $|\mu|\ge 2$.
\eqn\itfol{
\left|
|z_2|^2 + 2\bmu z_2 + |\mu|^2 + i (w +\zeta) 
\right|  \ge
\left|
|\mu|^2 + 2 \Re (\bmu z_2) + |z_2|^2 
\right|
= |\mu + z_2|^2.
}
Thus,
$$
|z_2'| \le {1\over {|z_2 + \mu|}} \le {1\over {|\mu|-|z_2|}} < 1.
$$
Also let us denote by,
$$
x = w + \zeta + 2 \Im (\bmu z_2).
$$
Then,
\eqn\wbnd{
|w'| = 
{{|x|}\over {\left| |z_2|^2 + 2\Re \bmu z_2 + |\mu|^2  + i x\right|^2}}
= {{|x|}\over {|z_2 + \mu|^2 + |x|^2}} < {{|x|}\over {1 + |x|^2}}
\le {1\over 2},
}
since $|z_2 + \mu| > 1$.
To complete the proof, we need to check the case $\mu = 0$.
In this case we know that $|\zeta|\ge 8$ and the same conclusion follows.

Now we can prove that all the terms in \genericg\ are different from
unity.
For this purpose we start with the point $(z_1=\infty, z_2)$ such
that $z_2$ is finite.
This is the classical region of the moduli space.
Acting either with $S$ or $S g(k_r,l_r,m_r)$ will turn that point into
$(0,0)$. The next $S g(k,l,m)$ which will act on it, will bring
it into the region \zbnd. The succesive $S g(k_i, l_i, m_i)$ will
not be able to take it out of \zbnd.
If we have an extra final $S$ (as in $h_2$ and $h_3$ of \genericg)
it will not bring it to the point at infinity because $|z_2| > 0$.


\listrefs
\bye
\end